\documentclass[preprint]{ptephy_v2}

\preprintnumber{XXXX-XXXX} 
\usepackage{hyperref}

\begin{document}

\title{Measurement of the isoscalar giant monopole resonance in $^{86}$Kr via deuteron inelastic scattering using an active target CAT-M}

\author[1,2,3*]{F.~Endo}
\author[1,4]{S.~Ota}
\author[4,5]{M.~Dozono}
\author[4]{R.~Kojima}
\author[1,6]{J.~Cai}
\author[7]{S.~Fracassetti}
\author[4]{S.~Hanai}
\author[8]{T.~Harada}
\author[4]{S.~Hayakawa}
\author[3,5]{Y.~Hijikata}
\author[4]{N.~Imai}
\author[3]{T.~Isobe}
\author[1,4]{K.~Kawata}
\author[4]{J.~Li}
\author[4]{S.~Michimasa}
\author[7]{R.~Raabe}
\author[4]{A.~Sakaue}
\author[4]{S.~Shimoura}
\author[3]{D.~Suzuki}
\author[9]{E.~Takada}
\author[3]{T.~Uesaka}
\author[4]{R.~Yokoyama}
\author[3,5]{J.~Zenihiro}
\author[6]{N.~Zhang}

\affil[1]{Research Center for Nuclear Physics, the University of Osaka, Ibaraki, Osaka 567-0047, Japan \email{fendo@rcnp.osaka-u.ac.jp}}
\affil[2]{Department of Physics, Tohoku University, Sendai 980-8578, Japan}
\affil[3]{RIKEN Nishina Center for Accelerator-Based Science, Wako, Saitama 351-0198, Japan}
\affil[4]{Center for Nuclear Study, the University of Tokyo, Wako, Saitama 351-0198, Japan}
\affil[5]{Department of Physics, Kyoto University, Kyoto, Kyoto 606-8502, Japan}
\affil[6]{Institute of Modern Physics, Chinese Academy of Sciences, Lanzhou, China}  
\affil[7]{Instituut voor Kern- en Stralingsfysica, K. U. Leuven, B-3001 Leuven, Belgium}
\affil[8]{Department of Physics, Toho University, Funabashi, Chiba 274-8510, Japan}
\affil[9]{National Institutes for Quantum and Radiological Science and Technology, Inage, Chiba, Japan}

\begin{abstract}%
Deuteron inelastic scattering on $^{86}$Kr was measured in inverse kinematics with the gaseous active target CAT-M, as part of a systematic investigation aimed at determining the nuclear matter incompressibility.
The isoscalar monopole strength distribution was extracted via multipole decomposition analysis, and the energy of the isoscalar giant monopole resonance was determined to be $17 \pm 1$~MeV.
The nuclear incompressibility of $^{86}$Kr and the isospin-dependent term of the nuclear matter incompressibility are discussed.
\end{abstract}

\subjectindex{xxxx, xxx}

\maketitle

\section{Introduction}\label{sec-intro}

The equation of state for nuclear matter is a fundamental equation that governs nucleon many-body systems, including nuclei and neutron stars.
It determines the properties of these systems and contributes to our understanding of astrophysical phenomena such as the mass-radius relationship of neutron stars and the mergers of massive stars.
In the field of experimental nuclear physics, efforts are being made to determine the equation of state near the saturation density and relatively small isospin asymmetry, from reactions and structures, such as mass, skin, and incompressibility \cite{ref-eos001}.
The incompressibility $K_0$ and its isospin-dependent term $K_\tau$ are sensitive to the nuclear matter equation of state under conditions of highly dense and large asymmetry \cite{ref-eos003}.
In a recent study, the impact of the uncertainty in the measurement of incompressibility on the equation of state has been investigated, suggesting that high-precision and high-accuracy measurements of the $K_\tau$ are still required for determining the equation of state \cite{ref-eos002}.

The isoscalar giant monopole resonance (ISGMR) is one of the collective motions of the nucleus, often referred to as the breathing mode, in which the protons and neutrons oscillate in the same phase and only in the radial direction.
The resonance energy and the nuclear incompressibility $K_A$ are related through $E_{\mathrm{ISGMR}} = \sqrt{ \hbar^2 A K_A / m \langle r^2 \rangle }$ \cite{ref-isgmr001}.
Here $E_{\mathrm{ISGMR}}$ is determined from the moment ratios ($\sqrt{m_3/m_1}$, where $m_n$ is the n-th energy-weighted moment of the strength function) of the ISGMR spectrum as defined in Eq.~\ref{eq-moment001}.
The ISGMR spectrum itself is extracted using the multipole decomposition analysis (MDA).
The incompressibility $K_A$ can be expressed based on the droplet model as follows \cite{ref-isgmr001}:
\begin{equation} \label{ka-dropmodel}
    K_A = K_\mathrm{0} (1 + cA^{-1/3}) + K_{\tau} \left( \frac{N - Z}{A} \right)^2 + K_\mathrm{C} Z^{2} A^{-4/3}
\end{equation}
 where $K_\mathrm{C}$ represents the Coulomb term, $K_\tau$ the asymmetry term, and $cA^{-1/3}$ the surface term in relation to the volume term.
The parameter $c$ accounts for finite-size effects of the nucleus and is assumed to be $-1.0$ \cite{ref-isgmr005} in the previous study\cite{ref-isgmr002}.
It has also been reported in previous studies that the theoretical uncertainty of $K_\mathrm{C}$ is small \cite{ref-isgmr004}.
By subtracting the Coulomb term from the nuclear incompressibility, $K_A$ can be expanded into zeroth- and second-order terms of the isospin asymmetry $(N-Z)/A$, where the coefficient of the second-order term corresponds to $K_\tau$.

Previous studies have measured the ISGMRs in stable nuclei such as Pb, Zr, and Sn, and the current values are reported as $K_0 = 240 \pm 20$ MeV and $K_\tau = -550 \pm 100$ MeV \cite{ref-isgmr002}.
However, the experimentally accessible range of the asymmetry has been limited to stable isotopes, and due to the assumption about the surface effect based on the liquid drop model, there remains a significant uncertainty in $K_\tau$.
Therefore, to determine $K_\tau$ with high precision and accuracy, it is necessary to systematically measure the ISGMR, including in unstable nuclei.

\begin{figure}[th]
    \begin{minipage}[b]{0.5\linewidth}
        \centering
        \includegraphics[keepaspectratio, scale=0.335]{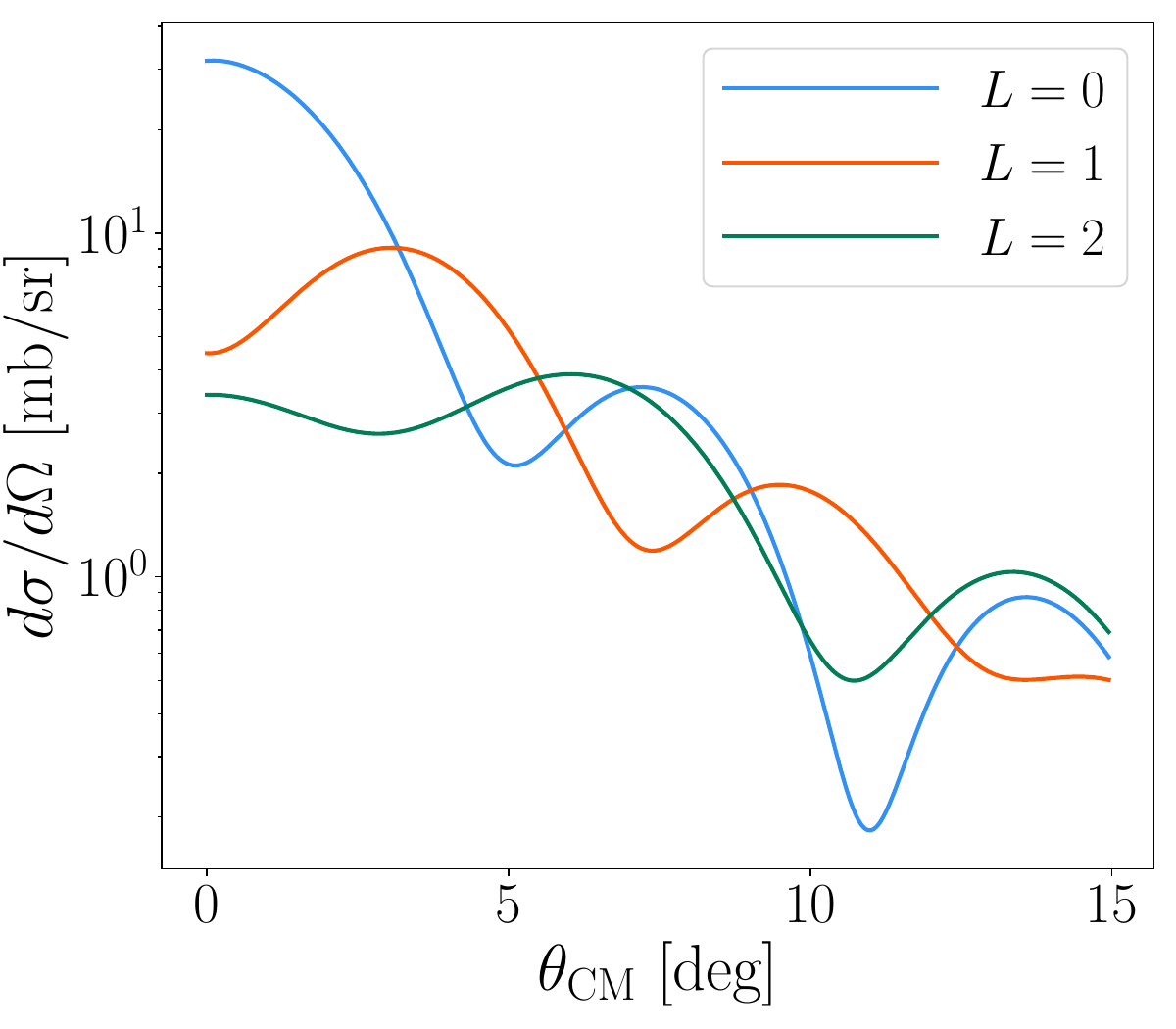}
    \end{minipage}
    \begin{minipage}[b]{0.5\linewidth}
        \centering
        \includegraphics[keepaspectratio, scale=0.335]{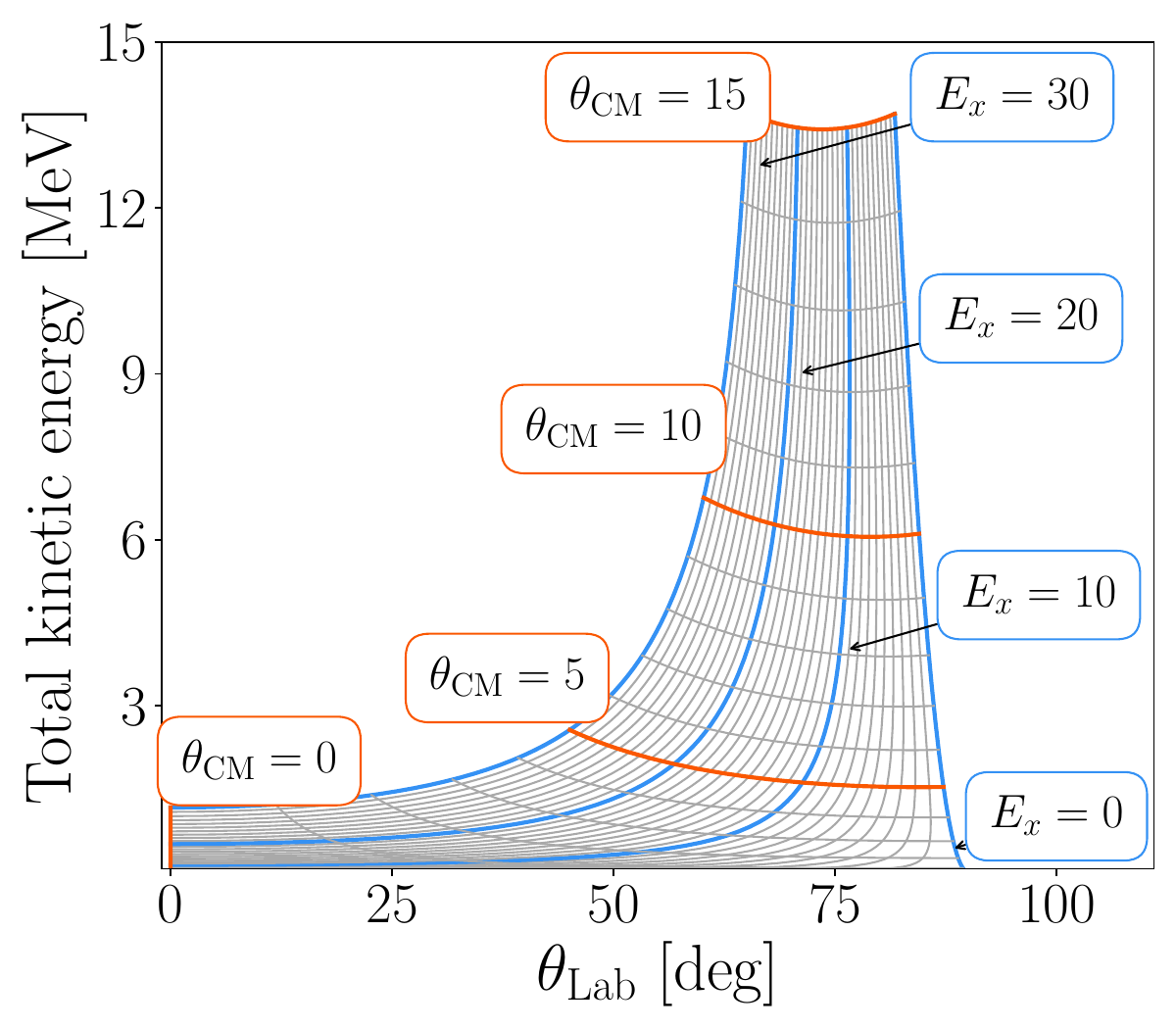}
    \end{minipage}
\caption{Left: Calculated angular distributions of differential cross section $d\sigma/d\Omega$ as a function of CM angle ($\theta_{\mathrm{CM}}$) for angular momentum transfers $\Delta L = 0,\, 1,\, 2$, obtained using the ECIS with a collective form factor for the $^{86}\mathrm{Kr}(d,d')$ at $E_x=21$~MeV and $100$~MeV/u incident energy. Right: Kinematics for the deuteron inelastic scattering of 100 MeV/u $^{86}\mathrm{Kr}$, showing the total kinetic energy and recoil angle in the laboratory system. Vertical curves represent correlations at constant excitation energy $E_x$ in MeV, while horizontal curves correspond to constant CM angles $\theta_\mathrm{CM}$ in degrees.}
\label{fig-intro-001}
\end{figure}

The ISGMR in unstable isotopes provides especially valuable information.
However, experimental measurements for such nuclei have long been technically challenging.
Figure \ref{fig-intro-001} illustrates the experimental difficulty.
The left panel displays calculated angular distributions for different angular momentum transfers, finding that the main component of $\Delta L = 0$ corresponds to the forward-angles. 
The right panel shows the typical kinematic correlation for recoil particles between the total kinetic energy and scattering angle in the laboratory system.
For forward-angle scattering, which is essential to extract the monopole component ($\Delta L = 0$) of the strength distribution, the corresponding recoil particles have very low energies—typically a few hundred~keV.
To measure these low-energy recoils, a thin target is required. On the other hand, thin targets reduce statistical significance. This presents a fundamental trade-off between detecting forward-scattered events and ensuring sufficient event statistics.
This technical limitation had been a significant barrier to ISGMR studies in unstable nuclei. With the development of active targets —serving simultaneously as both target and detector—this trade-off can now be effectively addressed.
As a pioneering previous study, ISGMR in $^{68}$Ni was conducted using the MAYA active target with two reactions: ($\alpha$, $\alpha’$) and ($d$, $d’$) \cite{ref-maya001, ref-maya002}.
However, in the case of inelastic $\alpha$ scattering, forward scattering events at angles less than $4$ degrees could not be measured due to the large energy loss of low-energy recoil particles.
In the case of deuteron inelastic scattering, break-up protons could not be excluded from the experimental data due to significant background.
Moreover, the statistics were not sufficient, and as a result, the strength function of the ISGMR could not be determined with sufficient precision due to large uncertainties in the multipole decomposition analysis.
An active target CAT-M [reference for CAT-M as "in- preparation" paper], developed for the measurement with high intensity beams more than 100~kcps, enables a wide angular coverage in the center-of-mass frame (approximately 2–10 degrees) by combining a low-pressure gaseous time projection chamber (TPC) capable of detecting recoil particles at forward angles, and silicon detectors for high-energy particles at backward angles.
Moreover, the installation of a dipole magnet also enhances the signal-to-noise ratio by rejecting delta electrons, produced by the heavy-ion beam interacting with the gas, from reaching the sensitive region of the Recoil TPC.
We performed an experiment to measure the ISGMR of $^{86}$Kr via deuteron inelastic scattering at Heavy Ion Medical Accelerator in Chiba (HIMAC) of the National Institute of Radiological Sciences (NIRS-QST) \cite{ref-himac001, ref-himac002, ref-himac003}.

In this paper, Section \ref{sec-experiment} describes the experimental setup and detector system. 
Section \ref{sec-analysis} details the data analysis procedures, and Section \ref{sec-result} presents the experimental results. In Section \ref{sec-discussion}, we discuss the systematic uncertainties and nuclear incompressibility in $N=50$ isotones.

\section{Experimental setup}\label{sec-experiment}
    
The ISGMR measurement of $^{86}$Kr was performed in the physics beam line (PH2 course) at HIMAC. 
The HIMAC accelerates heavy ions using two linear accelerators (RFQ and DTL) and a synchrotron with a $3.3$-second operation cycle.  
The instantaneous beam intensity was approximately $600$~kcps, with an average around $200$~kcps.
The energy of the $^{86}$Kr beam was $115$ MeV/u in front of the Al exit window of 0.1-mm thickness and the typical intensity was $2 \times 10^{5} $ particles per pulse.

\begin{figure}[th]
    \begin{minipage}[b]{0.275\linewidth}
        \centering
        \includegraphics[keepaspectratio, scale=0.225]{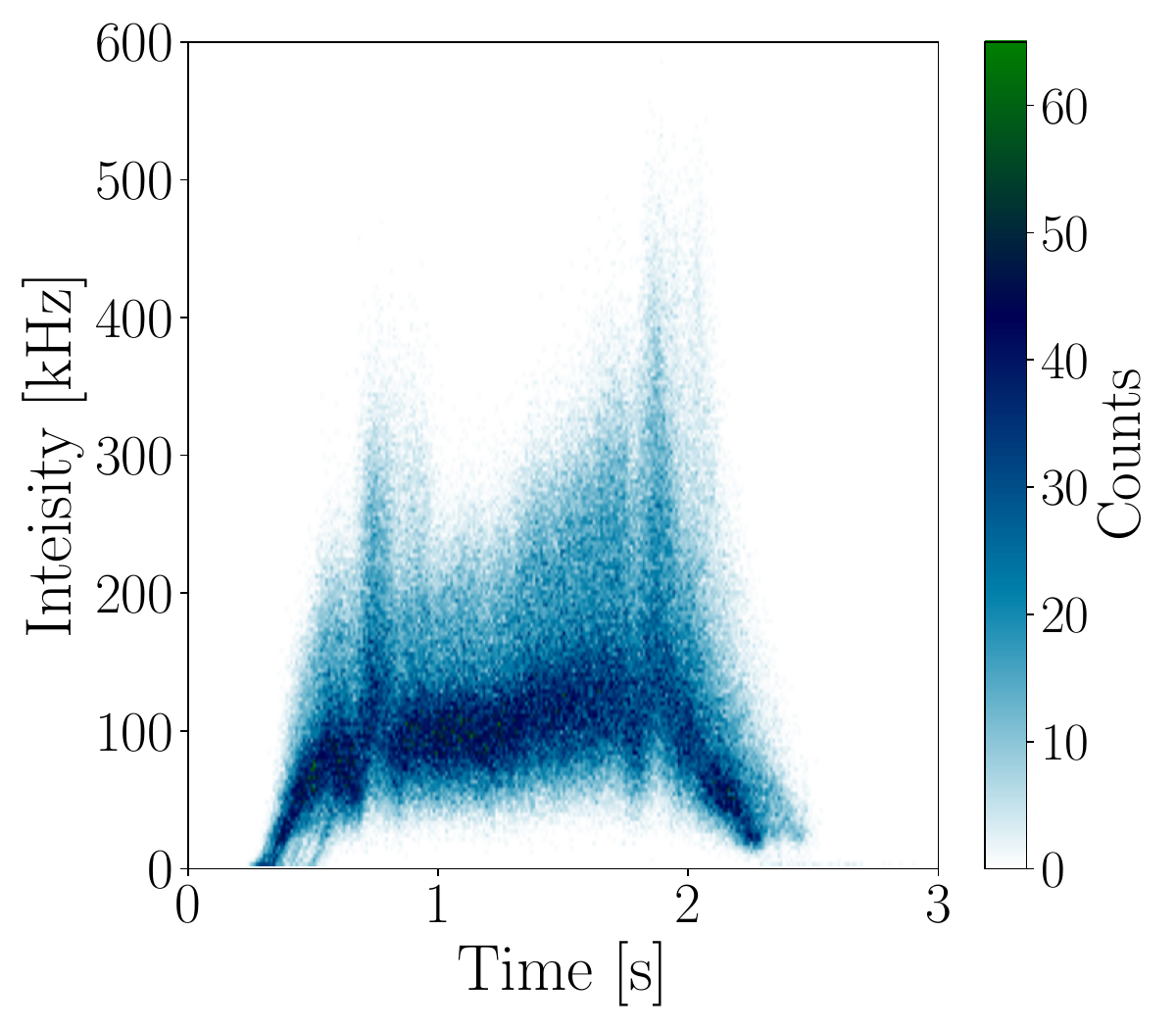}
    \end{minipage}
    \begin{minipage}[b]{0.725\linewidth}
        \centering
        \includegraphics[keepaspectratio, scale=0.1525]{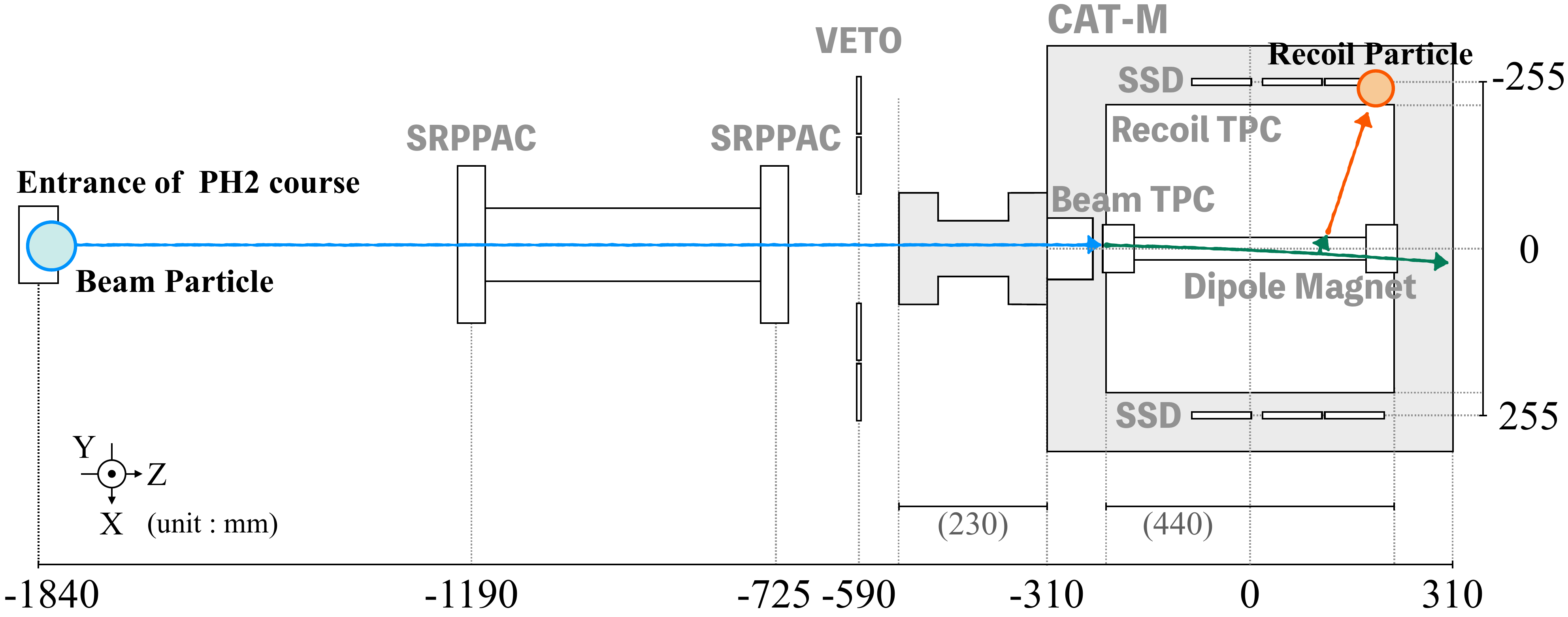}
    \end{minipage}
\caption{The left panel shows the spill structure of the incident beam, where the horizontal axis represents time and the vertical axis indicates beam intensity. The right panel shows the schematic view of the experimental setup. The beam travels from left to right. The two SR-PPACs, four plastic scintillators, and CAT-M are installed. The two SR-PPACs are connected by a single pipe filled with flowing isobutane gas at $10$~torr. Inside the CAT-M, the Beam TPC, Recoil TPC, SSDs, and a dipole magnet are installed. The CAT-M was filled with deuterium gas at $40$~kPa. The outside of the detectors was at atmospheric pressure, and it was separated from the internal gas by approximately $100$~$\mu$m thick Mylar films. The beam particles pass through the SR-PPACs and the Beam TPC, and then enter the dipole magnet. The recoil particles generated by the reaction in the magnet are affected by the magnetic field and subsequently detected by the Recoil TPC and SSDs.}
\label{fig-exp-001}
\end{figure}

As the beam position detector, two strip-readout parallel-plate avalanche counters (SR-PPACs) \cite{ref-srppac001,ref-srppac002}, four plastic scintillators for veto, and the CAT-M were installed from the upstream of the beamline.
Each SR-PPAC has one anode sandwiched by two cathodes with stripped electrodes, allowing it to determine the position by measuring charges accumulated on the strips.
It has an active area of $240 \times 150$ mm$^2$.
To remove beam particles that had undergone reactions or scattering upstream of CAT-M, four plastic scintillators were also installed as veto detectors.
The CAT-M, which has a chamber volume of $620 \times 480 \times 620$ mm$^3$, consists of a Beam TPC, a Recoil TPC, silicon strip detectors (SSDs), and a dipole magnet inside the Recoil TPC.
The Beam TPC has an active volume of $42 \times 12.1 \times 28$ mm$^3$, and employs two thick gas electron multipliers (THGEM) to multiply the ionization electrons. 
The 22 equilateral triangular pads with a side length of $5$ mm are used as readout pads.
The Recoil TPC has an active volume of $279 \times 200 \times 308$ mm$^3$.
A dual-gain multi-layer THGEM (DG-M-THGEM) and the 4048 equilateral triangular pads with a side length of $7$ mm are adopted as the amplification device and readout pads, respectively.
The dipole was installed inside the field cage of the Recoil TPC and generates a magnetic field over a volume of $28 \times 30 \times 320$~mm$^3$ using N42SH-R (Nd-Fe-B) permanent magnets and a return yoke.
The circular opening with a diameter of $30$~mm is provided at the entrance and exit of the magnet to allow the beam to pass through.
Twelve SSDs in total were arranged in a $3 \times 2$ configuration on each side of Recoil TPC,  and they have an active area of $92 \times 92$~mm$^{2}$, a thickness of $300$ $\mu$m with eight strips.
The deuterium gas at a pressure of $40$ kPa was circulated in CAT-M with the gas flow system, served as the active target. 
The dipole magnet is surrounded by the striped electrodes to maintain the uniform electric field in the TPC and the maximum deviation is $5$\% near the magnet according to the simulation.
The beam particle trajectories were measured by the SR-PPACs and  the Beam TPC, and the recoil particle trajectories were measured by the Recoil TPC and the SSDs.
The typical spill structure of the HIMAC and the experimental setup are shown in Figure \ref{fig-exp-001}.

For data acquisition, two systems were employed: the general electronics for TPCs (GET) \cite{ref-daq006} and a typical DAQ system at RIBF in RIKEN (babirl) \cite{ref-daq001}.
The GET system in this experiment consisted of 18 AsAds, five CoBos, and one MuTanT and was employed for the data acquisition of the CAT-M. 
The signals from the readout pad of the Recoil and Beam TPC and silicon detectors were input into AsAds through the ZAP board for discharge protection.
The signals from the cathode electrodes of the SR-PPAC were input into preamplifier-shaper-discriminator board RPA-132, product of HAYASHI-REPIC Co. Ltd., and the timing and width of output logic signals digitized by V1190 TDCs, product of CAEN Co. Ltd. 
The signals from the anode electrode of SR-PPAC were input into a preamplifier, fast amplifier and constant fraction discriminator. 
The logic signals were input into V1290 TDCs, products of CAEN Co. Ltd. The data from V1190 and V1290 TDCs were read out by MOCOs and the mountable controller with parallelized VME (MPV) \cite{ref-daq002, ref-daq005} in the babirl system.
Trigger signals generated at each readout electrode were first sent from the corresponding CoBo boards of the GET DAQ to the genetic trigger operator (GTO) \cite{ref-daq003, ref-daq004}. The GTO selected accepted triggers and returned them, along with busy signals, to the MuTanT module, thereby sharing the DAQ trigger logic.
The number of beam particles and DAQ efficiency were counted using a scaler (SIS3820), recording the anode signals from SR-PPAC, a $20$-MHz clock, and the number of clock counts when the GTO was in accepted state. 
The measured DAQ efficiency was $51$\%. 
Beam particle pile-up was corrected based on the total charge measured in the Beam TPC. 
The number of  target nuclei was estimated using a pressure gauge with a resolution of $0.1$ kPa, recorded by a data logger.

\section{Data analysis}\label{sec-analysis}

This section describes the method for deriving the double-differential cross section of deuteron inelastic scattering. 
The excitation energy and scattering angle in the center-of-mass system are deduced using the four-momenta of the beam and recoil particles before and after the vertex, based on the missing mass spectroscopy.

In the analysis, particle identification and tracking of the beam and recoil particles were performed to reconstruct the four-momentum at the reaction vertex. 
However, since the reaction point was located inside the dipole magnet, which was installed inside the field cage of Recoil TPC, the vertex was not measured.
Therefore, the trajectories were extrapolated into the magnetic field region, and the four-momentum at the vertex was reconstructed accordingly.
Finally, the double-differential cross section were estimated by correcting for the relevant efficiencies, including tracking and transmission efficiencies from the analysis, the DAQ efficiency, and the detector acceptance evaluated from Monte Carlo simulations.
Eq.~\ref{eq-def-dsdedo} shows the components contributing to the double-differential cross section in this analysis process.
\begin{equation}
    \frac{d^2\sigma}{dEd\Omega} \left( \theta_{\mathrm{CM}},\,E_x\right)
    \equiv
    \frac{
    \mathrm{Yield}\left( \theta_{\mathrm{CM}},\,E_x\right)
    }
    { 
    N_{\mathrm{beam}} N_{\mathrm{target}} 
    \epsilon_{\mathrm{beam}} \epsilon_{\mathrm{DAQ}} \epsilon_{\mathrm{recoil}} \epsilon_{\mathrm{trans}} 
    \epsilon_{\mathrm{sim}} \left( \theta_{\mathrm{CM}},\,E_x\right)
    \Delta E \Delta \Omega \left( \theta_{\mathrm{CM}}\right) 
    }, \label{eq-def-dsdedo}
\end{equation}
where \(N_{\mathrm{beam}}\) and \(N_{\mathrm{target}}\) denote the number of beam and target particles, respectively. 
The \(\epsilon_{\mathrm{beam}}\) and \(\epsilon_{\mathrm{trans}}\) correspond to the tracking efficiency of beam particles and the overall transmission efficiency from the entrance and exit of the dipole magnet, respectively. 
The $\epsilon_{\mathrm{recoil}}$ is also the efficiency of tracking and identifying recoil particles using the Recoil TPC and/or SSD.
The \(\epsilon_{\mathrm{DAQ}}\) and \(\epsilon_{\mathrm{sim}}\) represent the data acquisition efficiency and the acceptance of recoil particles obtained from Monte Carlo simulation.
The yield \( \mathrm{Yield}\left( \theta_{\mathrm{CM}},\,E_x \right) \) was reconstructed from events obtained by missing mass spectroscopy.
A summary of all variables and correction factors used in the cross section calculation is provided in Table~\ref{table-dsdode-summary}.
 \begin{table}[th]
 \caption{Summary of the variables of the  cross section. The detailed analysis of each variable is described in the right-hand column.}
 \label{table-dsdode-summary}
 \centering
  \begin{tabular}{c|l|c|c}
   \hline
   Variables & Description & Detail  & Typical values\\
   \hline \hline
   $\mathrm{Yield}\left( \theta_{\mathrm{CM}},\,E_x\right)$          & Counts obtained from data analysis.       & Section~\ref{sub-ana-vertex}                              &  - \\
   $N_{\mathrm{beam}}$                                               & Number of beam particle.                  & Section~\ref{sec-experiment}, \ref{sub-ana-beam-track}    &  - \\
   $N_{\mathrm{target}}$                                             & Number of target particle.                & Section~\ref{sec-experiment}                              &  - \\
   $\epsilon_{\mathrm{beam}}$                                        & Tracking efficiency of beam particle.     & Section~\ref{sub-ana-beam-track}                          &  $99$\%\\
   $\epsilon_{\mathrm{trans}}$                                       & Beam transmission.                        & Section~\ref{sub-ana-acceptance}                          &  $91$\% \\
   $\epsilon_{\mathrm{DAQ}}$                                         & DAQ efficiency.                           & Section~\ref{sec-experiment}                              &  $51$\% \\
   $\epsilon_{\mathrm{recoil}}$                                      & Tracking efficiency of recoil particle.   & Section~\ref{sub-ana-recoil-track}                        &  $87$\% or $98$\% \\
   $\epsilon_{\mathrm{sim}}\left( \theta_{\mathrm{CM}},\,E_x\right)$ & Detection efficiency of recoil particles. & Section~\ref{sub-ana-acceptance}                          &  $<27$\% \\
   \hline
  \end{tabular}
\end{table}

    \subsection{4-momentum for recoil particle}\label{sub-ana-recoil-track}

    The particle identification and tracking were performed using a combination of the Recoil TPC and SSD when the particle reached the SSD.
    Prior to tracking, signals recorded on the readout pads of the Recoil TPC were clustered using a single-linkage method \cite{ref-cluster001}, which grouped spatially connected pads. This procedure served to remove noise and eliminate events with multiple recoil particles arriving simultaneously.    
    The trajectory was determined by fitting the positions and measured charges from the Recoil TPC readout pads and the SSD strips using stopping powers for protons and deuterons. The effect of diffusion of drift electrons was taken into account during the fitting process. The stopping powers were calculated by SRIM \cite{ref-srim001}.
    
    In this experiment, recoil particles either stopped within the Recoil TPC or reached the SSD.
    For those that stopped inside the Recoil TPC, particle identification and trajectory reconstruction were performed simultaneously, based on the stopping powers. In contrast, for recoil particles that reached the SSD, particle identification was first conducted using the so-called $\Delta E$–$E$ method based on the energy loss from both the Recoil TPC and the SSD, followed by tracking using the same procedure.
    \begin{figure}[th]
        \begin{minipage}[b]{0.5\linewidth}
            \centering
            \includegraphics[keepaspectratio, scale=0.35]{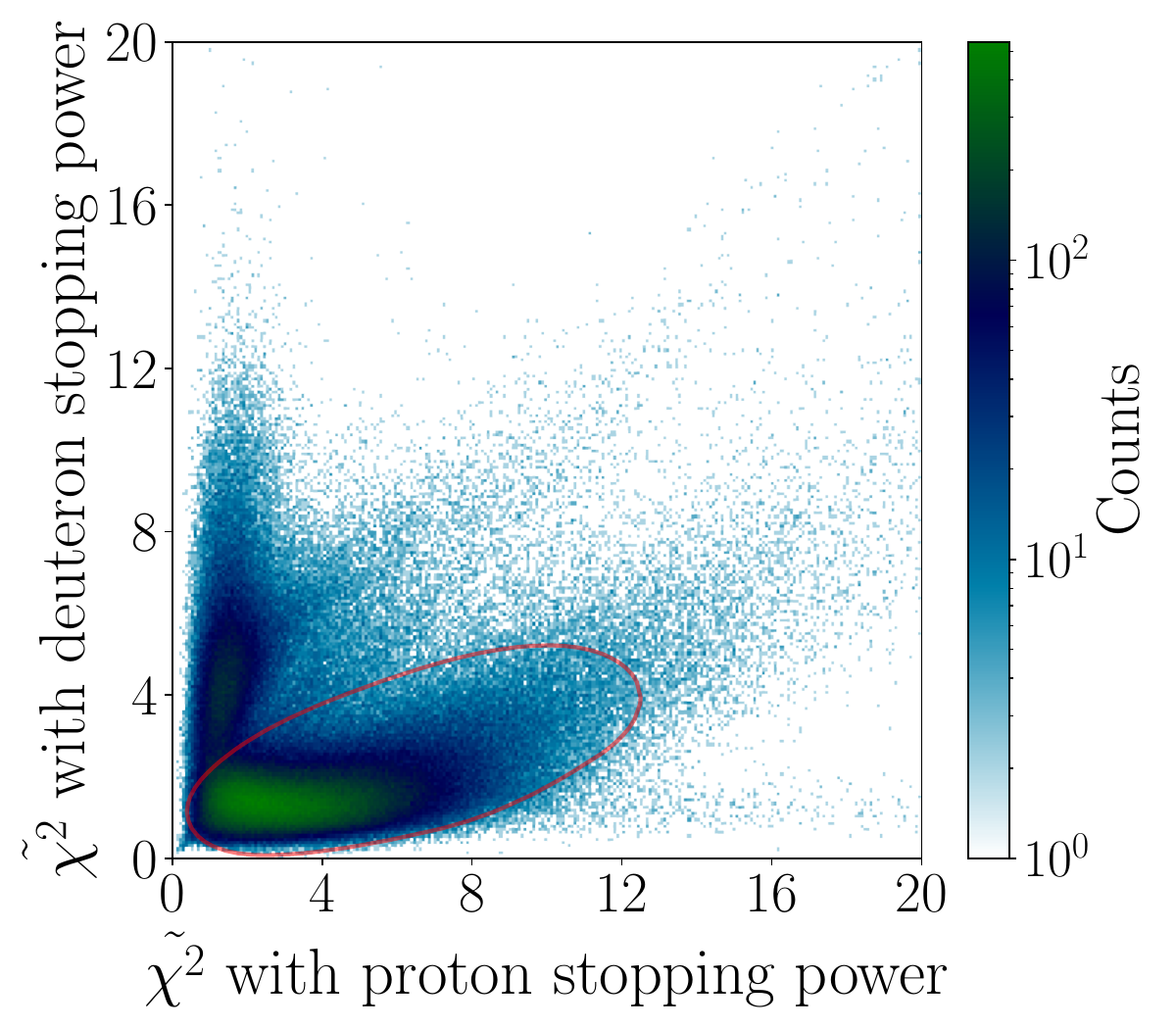}
        \end{minipage}
        \begin{minipage}[b]{0.5\linewidth}
            \centering
            \includegraphics[keepaspectratio, scale=0.35]{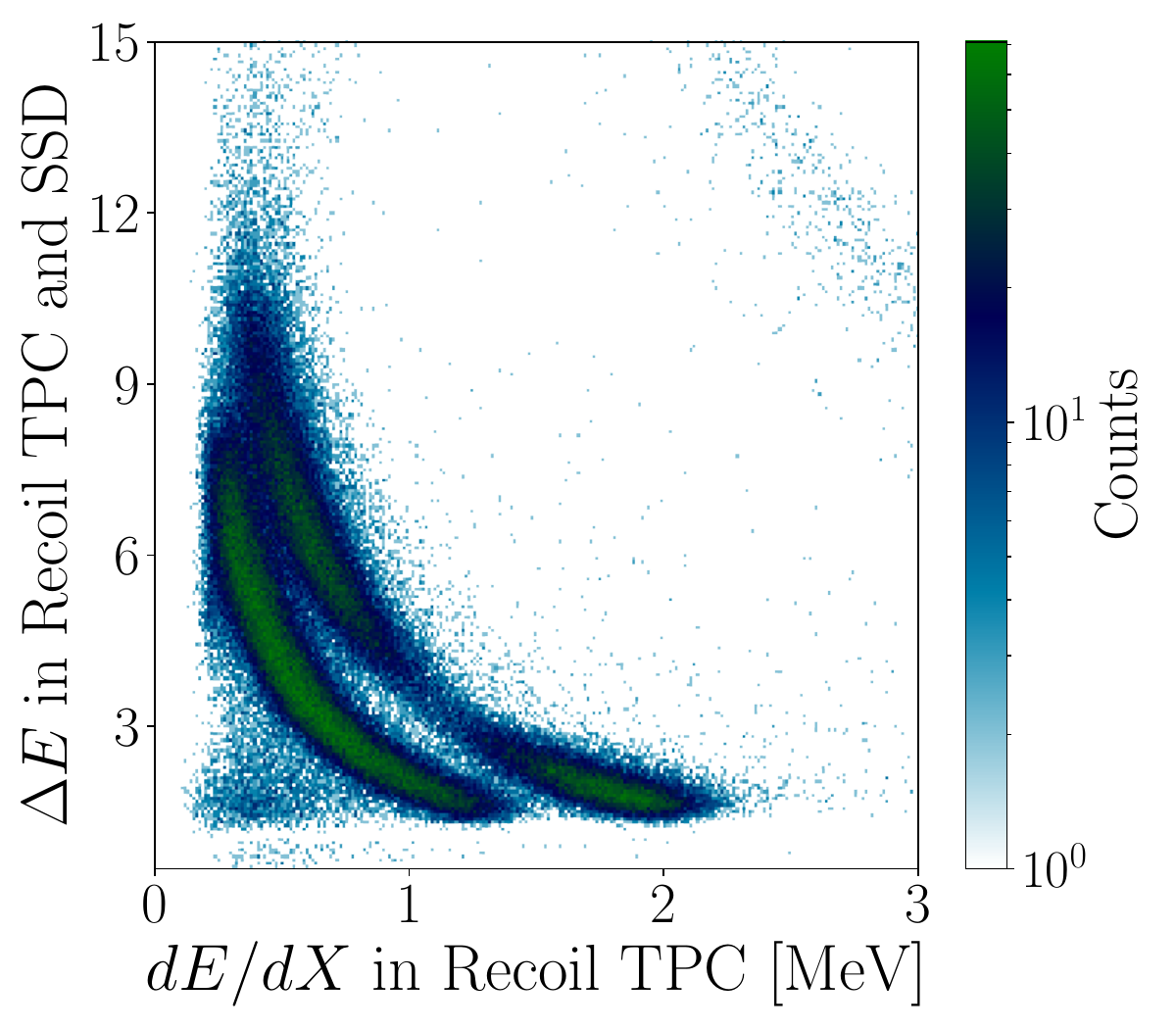}
        \end{minipage}
    \caption{Particle identification of recoil particles. The left panel shows the chi-squared correlations of the tracking results using proton and deuteron stopping power. This method is used when the recoil particles stop inside the Recoil TPC. The right panel is the $\Delta E$-$E$ correlation using the energies measured at the Recoil TPC and SSD. It is used when the recoil particles reach the Recoil SSD.}
    \label{fig-recoil-pi-001}
    \end{figure}
    Figure~\ref{fig-recoil-pi-001} (left) shows the result of particle identification for particles that stopped in the Recoil TPC. It presents the correlation of reduced $\chi^2$ values obtained from tracking with the stopping powers of protons and deuterons. The region enclosed by the red line corresponds to deuterons, and the blue region corresponds to protons. This particle identification method exploits the difference in stopping power near the Bragg peak for each particle.
    Figure~\ref{fig-recoil-pi-001} (right) shows the $\Delta E$–$E$ correlation, where the vertical axis represents the total energy loss in the Recoil TPC and SSD, and the horizontal axis indicates the average energy loss in the Recoil TPC. The upper locus corresponds to deuterons, while the lower one corresponds to protons. 
    
    As a result of this tracking analysis, the tracking efficiency was found to be $98$\% for particles stopping in the Recoil TPC, and $87$\% for those reaching the SSD, with systematic uncertainties of $5$\% and $3$\%, respectively.

    \subsection{4-momentum for beam particle}\label{sub-ana-beam-track}
    
\begin{figure}[th]
        \begin{minipage}[b]{0.5\linewidth}
            \centering
            \includegraphics[keepaspectratio, scale=0.35]{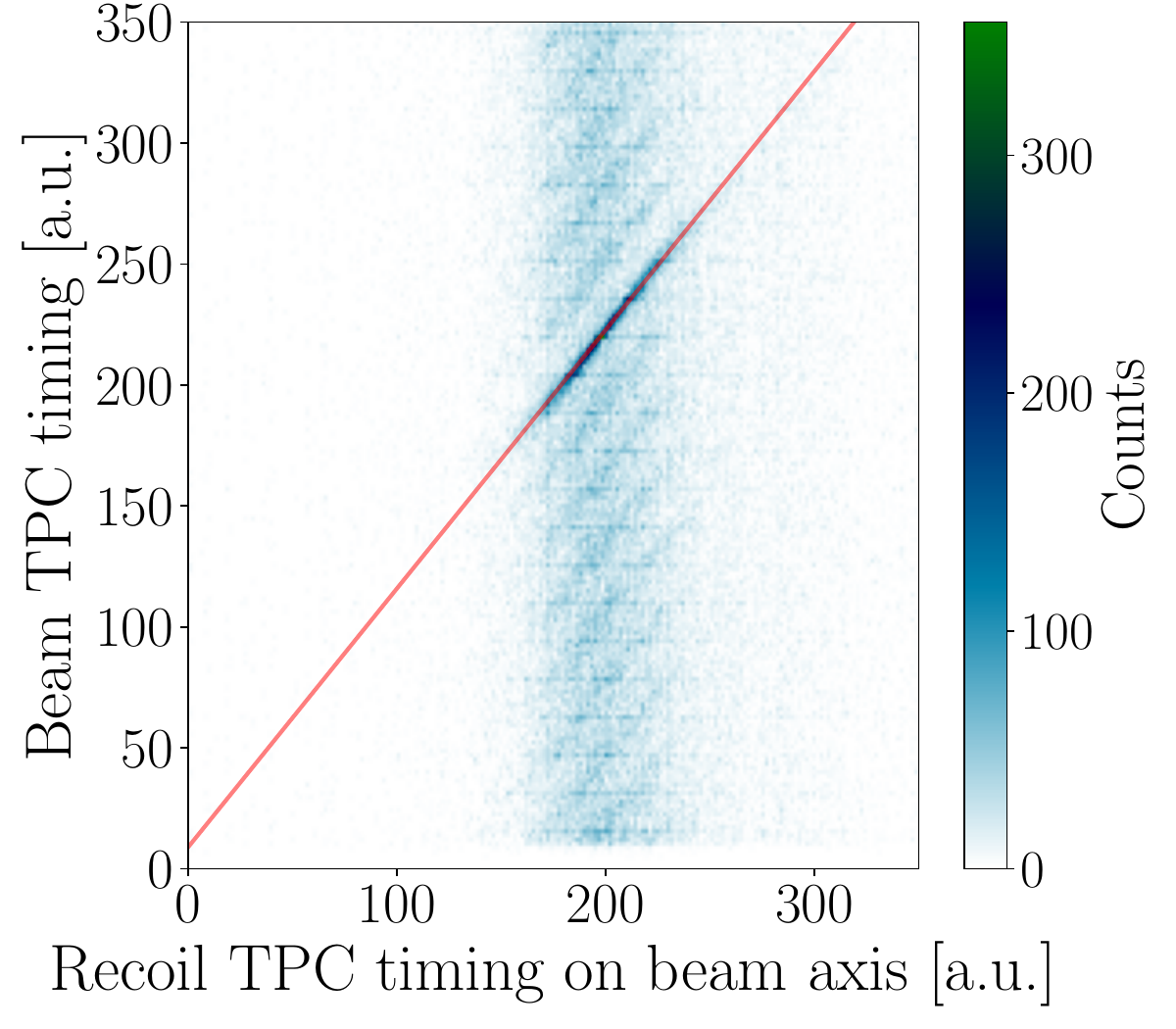}
        \end{minipage}
        \begin{minipage}[b]{0.5\linewidth}
            \centering
            \includegraphics[keepaspectratio, scale=0.35]{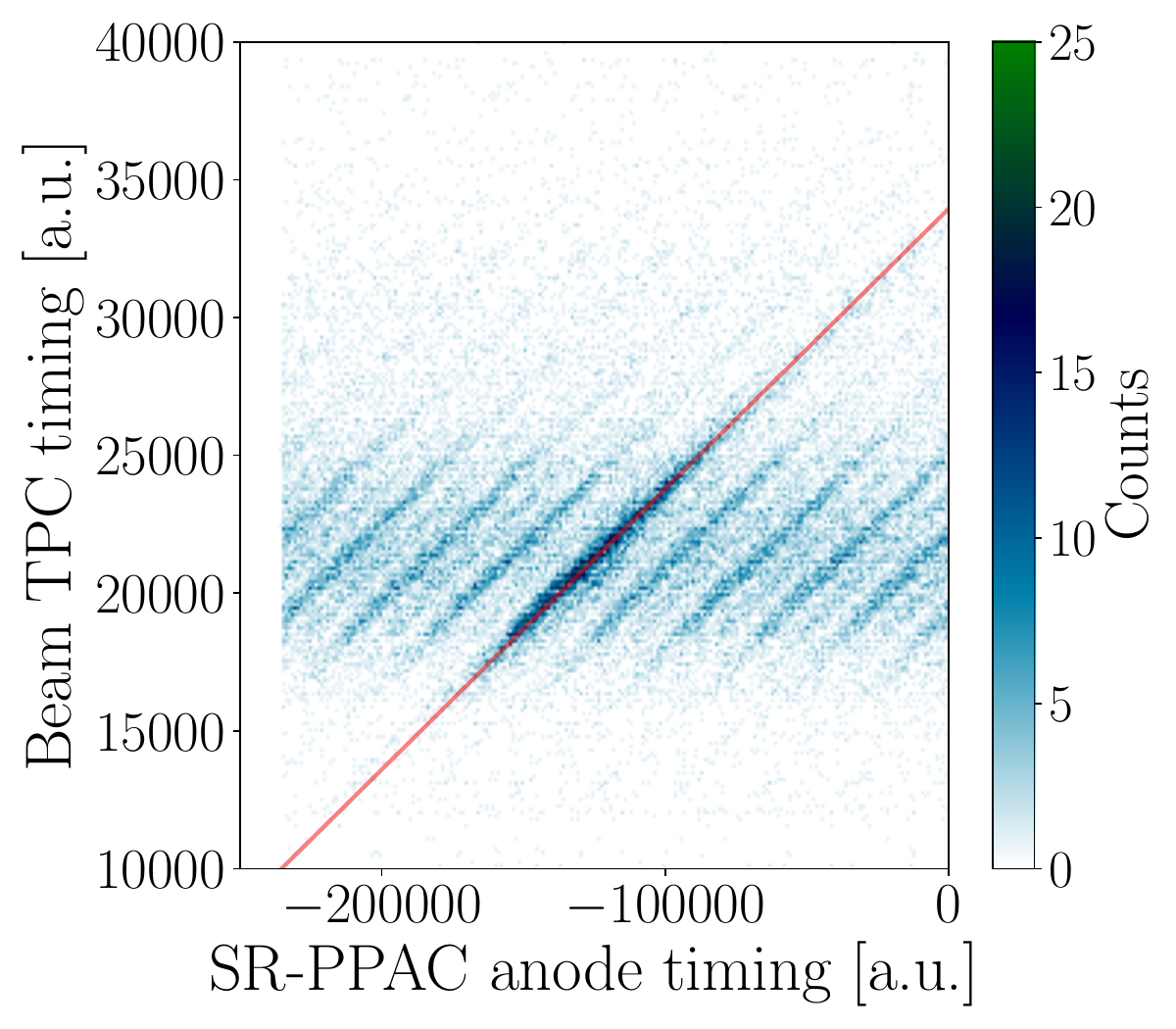}
        \end{minipage}
    \caption{Left: Timing correlation between the Beam TPC and the Recoil TPC. The locus, which has the most counts, corresponds to beam particles that undergo a reaction. A linear fit was applied to the trend of the mode of this locus, and events with a shortest distance to the fit below $10$ were identified as reaction-related beam particles. Right: Timing correlation between the Beam TPC and the anode of upstream SR-PPAC. Event selection was performed in the same manner for all eight SR-PPAC planes (anode and cathode for each SR-PPAC) as in the case of Beam TPC. The distance thresholds for event identification were set to $1000$ for anodes and $800$ for cathodes.}
    \label{fig-beam-gate}
    \end{figure}

    The beam particle trajectory was determined using positions measured by two SR-PPACs and the Beam TPC.
    Before determining the positions in each detector, events scattered with the recoil particle were selected based on the timing correlation with the clustering results of the Recoil TPC.
    Figure~\ref{fig-beam-gate} (left) shows the timing correlation between the Beam TPC and the Recoil TPC.
    The horizontal axis represents the timing on the beam axis extrapolated from the clustering result of the recoil particle, and the vertical axis represents the timing measured at each pad of the Beam TPC.
    The locus along the red line in the figure corresponds to the beam particles that reacted with the recoil particles.
    In this analysis, events near this red line were selected.
    
    On the other hand, Figure~\ref{fig-beam-gate} (right) shows the timing correlation between the Beam TPC and the anode of the SR-PPAC, which is installed at the most upstream position.
    The horizontal axis represents the timing measured at the anode of the SR-PPAC, and the vertical axis shows the timing of valid events in the Beam TPC.
    For the SR-PPACs, events were similarly selected for all electrodes (both anodes and cathodes of the two SR-PPACs).
    
    The beam particle positions were determined for each detector after event selection, and these were combined to reconstruct the trajectories.
    In the Beam TPC, the position on the XZ plane was determined by weighting the pad positions with the measured charges, and the Y-position was derived from the drift time using the drift velocity estimated by Garfield++ \cite{ref-gafieldpp001}.
    For the SR-PPACs, the charge of each strip was determined using the time-over-threshold method, and the positions were calculated using the $Q_LQ_R$ and $Q_0Q_1$ methods \cite{ref-srppac001, ref-srppac002}.

\begin{figure}[th]
        \begin{minipage}[b]{0.5\linewidth}
            \centering
            \includegraphics[keepaspectratio, scale=0.35]{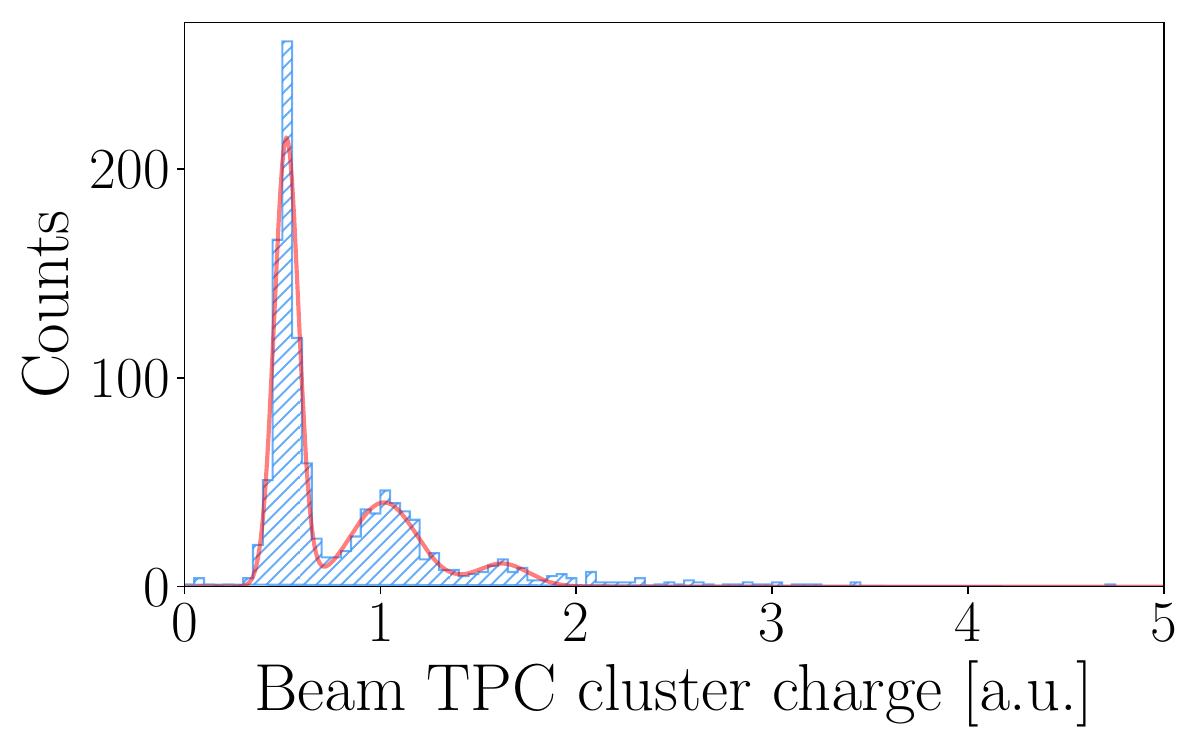}
        \end{minipage}
        \begin{minipage}[b]{0.5\linewidth}
            \centering
            \includegraphics[keepaspectratio, scale=0.35]{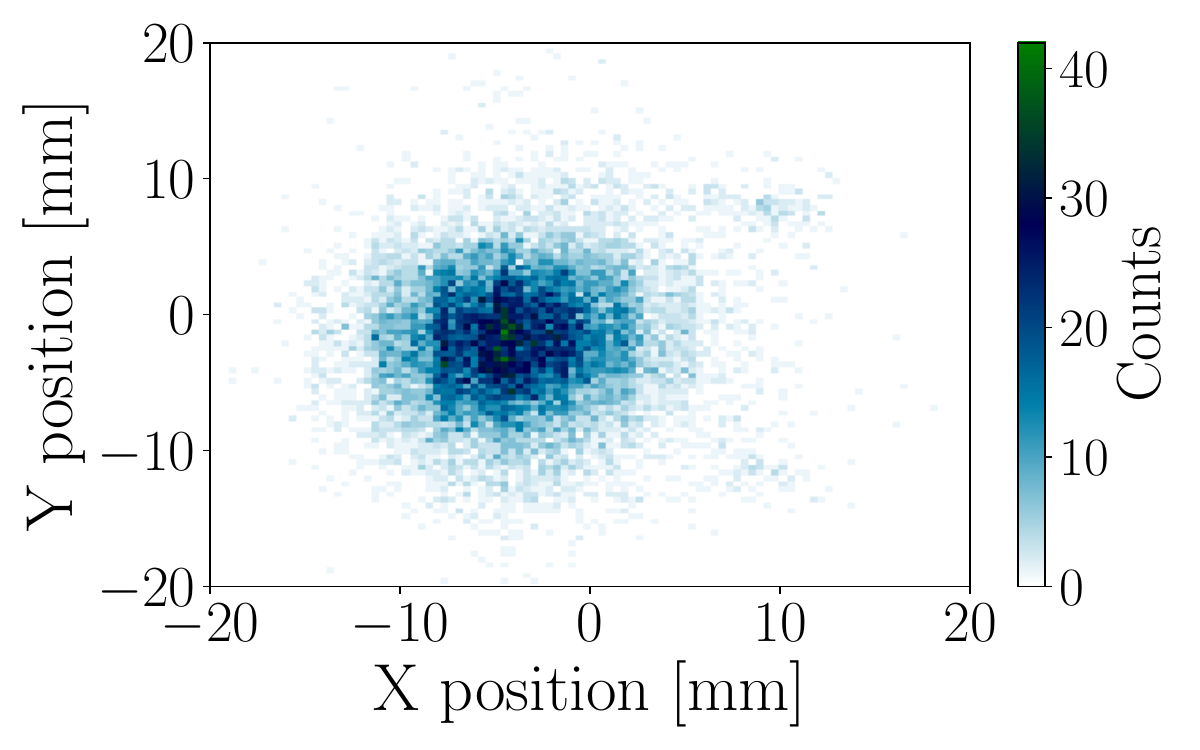}
        \end{minipage}
    \caption{Left: Total cluster charge measured on each pad of the Beam TPC after beam selection. Since the beam particles enter the Beam TPC at a negligibly small angle, periodic peak structures corresponding to the number of piled-up particles appear. In this analysis, the distribution was fitted using the sum of three Gaussian functions to estimate the average number of pile-up particles. Right: Extrapolated position for beam particle at the center of the Recoil TPC, obtained from tracking analysis. The tracking was performed using position information derived from the Beam TPC and SR-PPACs. The influence of the magnetic field was ignored in this extrapolation.}
    \label{fig-beam-track-001}
    \end{figure}

    Beam particle tracking was performed when position information was available from at least two of these detectors, and the trajectory was determined independently in the XZ and YZ planes.
    The tracking efficiency was defined as the ratio of successfully reconstructed tracks to the number of events where recoil particles were detected, yielding a result of approximately $99\%$.
    Figure \ref{fig-beam-track-001} (left) shows the beam profile at the center of the Recoil TPC, obtained by extrapolating the determined tracks.
    
    Furthermore, the mean pile-up count was estimated from the sum of measured charge at each pad after event selection (Figure \ref{fig-beam-track-001}).
    The distribution shows three equally spaced peaks, corresponding to single-, double-, and triple-particle events.
    Therefore, by fitting the distribution with three Gaussian functions and calculating the weighted mean of their integrals, the average pile-up rate was determined as $1.5$.
    The contribution from events with four or more particles was neglected in this estimation.
    The $N_{\mathrm{beam}}$ was derived from the product of the scaler value and the average pile-up rate.
    
    \subsection{Vertex Reconstruction}\label{sub-ana-vertex}

    The vertex and reaction were reconstructed using the beam and recoil tracks obtained through the methods described in previous subsections. 
    In this experiment, the reaction itself occurred inside the dipole magnet in CAT-M, therefore, it was necessary to correct the particle trajectories under the magnetic field before vertex reconstruction.
    
    For the beam particle, the total kinetic energy is high, and the energy loss within the CAT-M is small. Moreover, the beam particles passed near the center of the dipole magnet, where the variation in magnetic field strength was negligible. Therefore, the trajectory was corrected using the analytical solution to the equation of motion for a charged particle traveling at constant velocity in a uniform magnetic field.
    On the other hand, the total kinetic energy of the recoil particles ranges from approximately $500$~keV to $12$~MeV. In particular, when the total kinetic energy is around $500$~keV, the energy loss within the effective magnetic field region can reach approximately 50\% of the total kinetic energy.
    Additionally, the spatial variation of the magnetic field along the recoil emission direction is non-negligible. Therefore, assuming a uniform field or constant velocity was unrealistic. Thus, the recoil particle tracks were corrected by numerically solving the equations of motion for a slowing charged particle in a non-uniform magnetic field. The magnetic field was modeled using measured magnetic field map. Details of the field modeling and comparisons of the pre- and post-correction trajectories for both beam and recoil particles are described in Appendix~\ref{ape-magnet-model}.
    
    The reaction vertex was determined by finding the point of closest approach between the corrected beam and recoil tracks. Using the reconstructed four-momenta of the beam and recoil particles at the vertex, the excitation energy spectrum was obtained via missing mass spectroscopy.
    \begin{figure}[th]
        \begin{minipage}[b]{0.5\linewidth}
            \centering
            \includegraphics[keepaspectratio, scale=0.38]{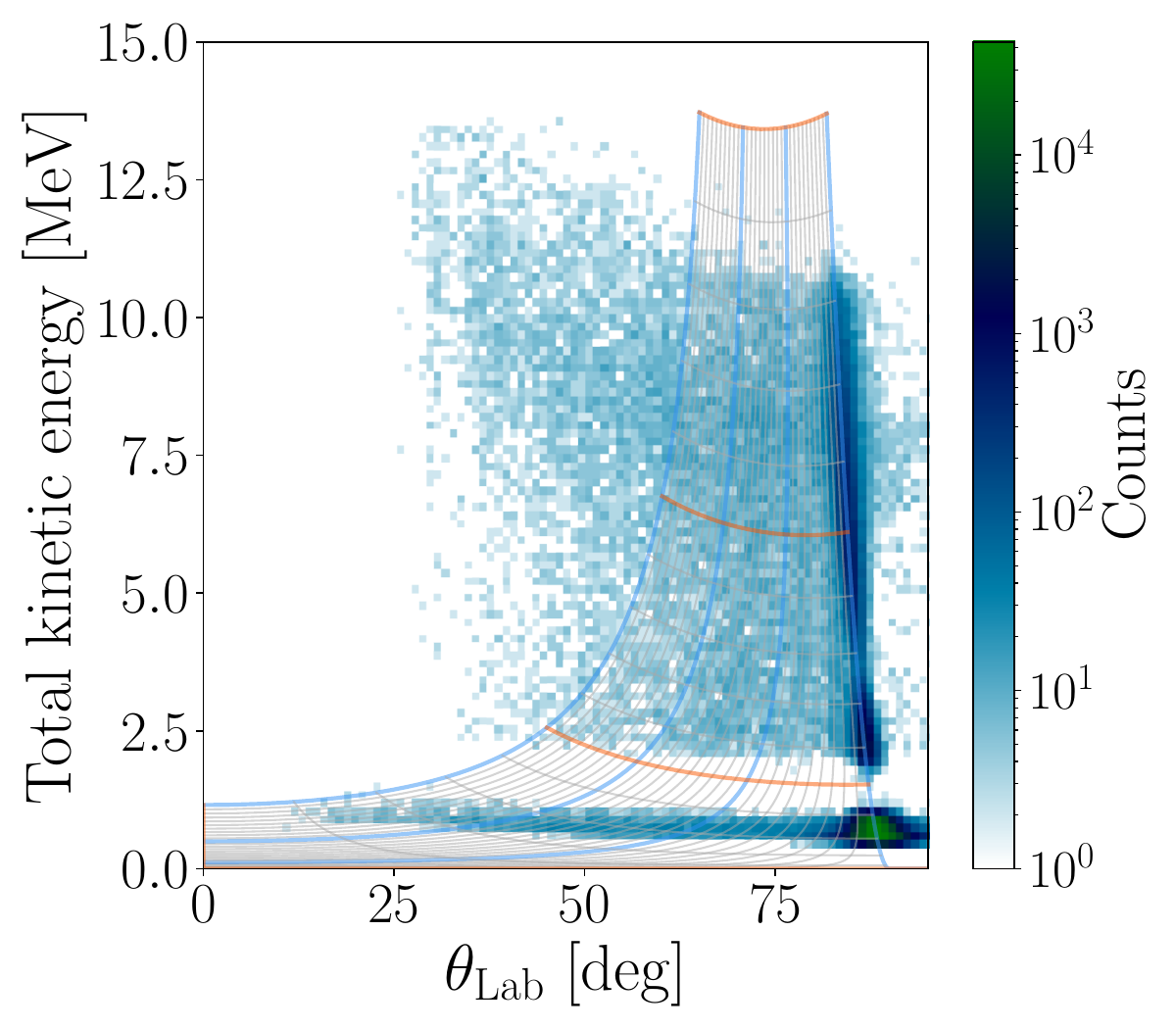}
        \end{minipage}
        \begin{minipage}[b]{0.5\linewidth}
            \centering
            \includegraphics[keepaspectratio, scale=0.38]{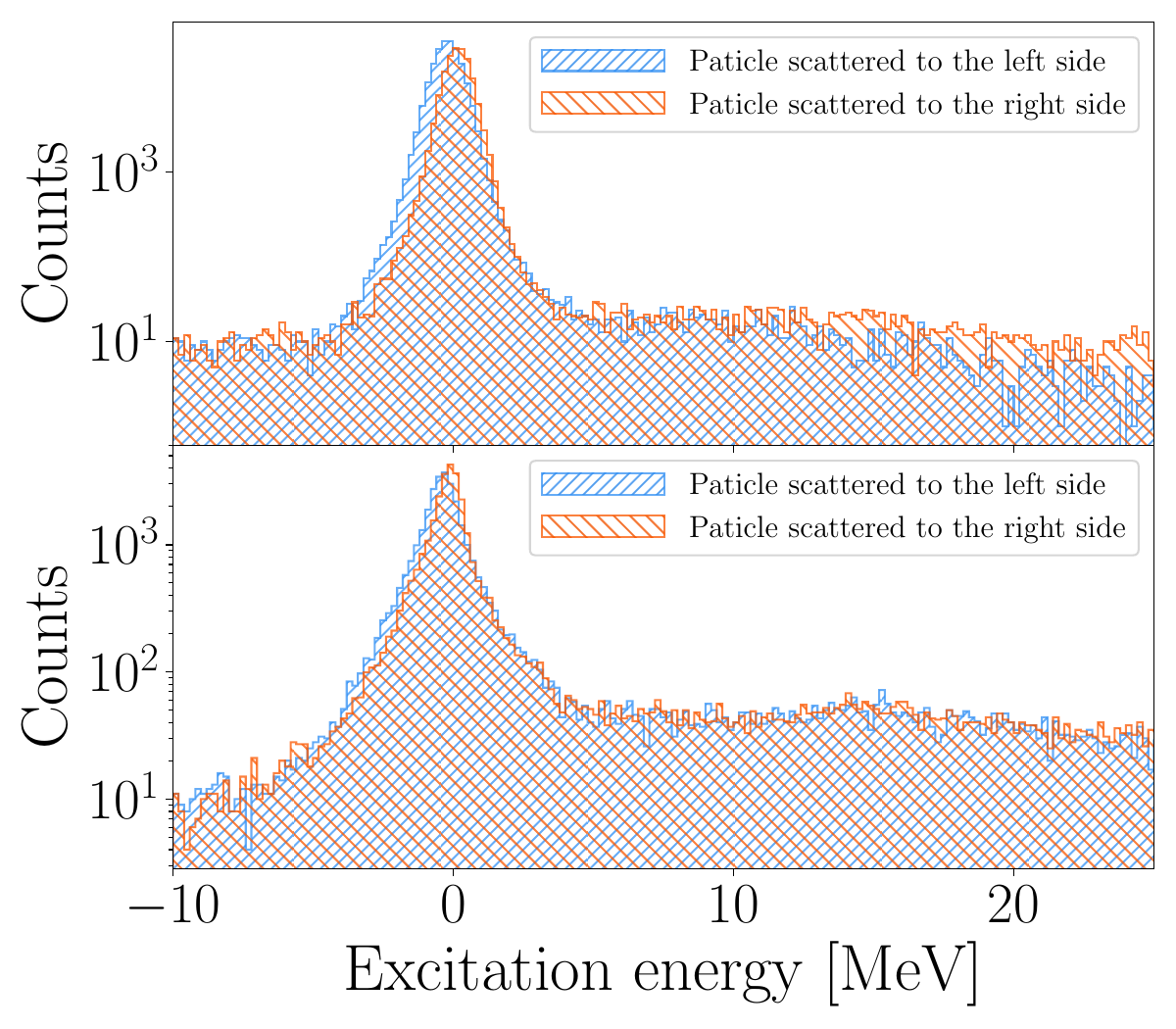}
        \end{minipage}
    \caption{Left: Kinematic correlation of recoil particles after vertex reconstruction. The color map represents experimental data, while the solid lines indicate the kinematic curves calculated based on reaction kinematics (see Figure~\ref{fig-intro-001} for details). The lower locus corresponds to events where the recoil particles stopped inside the Recoil TPC, and the upper locus corresponds to events where the recoil particles reached the SSD. Right: Excitation energy spectra. The top panel shows events in which recoil particles stopped in the Recoil TPC, while the bottom panel shows events that reached the SSD. In both panels, the blue hatched histogram represents events scattered to the left of the beam axis, and the orange hatched histogram represents those scattered to the right.}
    \label{fig-kinematics-ex}
    \end{figure}
    Figure~\ref{fig-kinematics-ex} shows the kinematic correlations of the recoil particles and the resulting excitation energy spectrum. The left panel displays the correlation between the total kinetic energy and the recoil angle in the center-of-mass system at the vertex. The locus near $500$~keV corresponds to particles that stopped inside the Recoil TPC, while the broader distribution above $2$~MeV corresponds to those reaching the SSD. As shown in Fig.~\ref{fig-exp-001}, the SSD is positioned approximately $35$~mm away from the field cage of the Recoil TPC, resulting in an insensitive region corresponding to an energy gap around $1.5$~MeV.
    The right panel of Fig.~\ref{fig-kinematics-ex} presents the excitation energy spectra: the upper histogram corresponds to particles stopped in the Recoil TPC, and the lower histogram to those reaching the SSD. Blue histograms represent particles scattered to the left side (viewed from the beam direction), while orange represents particles scattered to the right.
    The difference in the centroid of the elastic scattering peaks reconstructed from particles scattered to the left and right corresponds to the systematic uncertainty associated with the analysis method mentioned above. This results in a systematic uncertainty of approximately $360$~keV for recoil particles that stopped within the TPC, and about $200$~keV for those that reached the SSD.
    
    The typical bending angle of recoil particles during the effective magnetic field was also about $4$~degrees in the laboratory system, and the resulting difference in excitation energy between particles scattered to the left and right exceeded $4$~MeV without correction. However, our current correction reduced this shift to less than $400$~keV. The remaining discrepancy of approximately $400$~keV is primarily attributed to systematic uncertainties in magnetic field modeling. This was confirmed by comparing simulations based on GEANT4 (a software toolkit for simulating particle trajectories in matter \cite{ref-geant4001}) using the measured field map, with reconstructed trajectories based on the magnetic field model.
    
    \subsection{Acceptance}\label{sub-ana-acceptance}

    To derive the double-differential cross section, the transmission of the beam particles and the acceptance of the recoil particles was evaluated.
    For the beam particles, the beam profile at the entrance window of the dipole magnet was obtained from the trajectories before magnetic field correction, while the profile at the exit window of the dipole magnet was determined from the trajectories after correction.
    The transmission ($\epsilon_{\mathrm{trans}}$) was determined as the ratio of events that passed through both the entrance and exit windows ($30$~mm in diameter) of the dipole magnet to the total number of events that were tracked, yielding a value of $0.91$.
    For the recoil particles, the acceptance for recoil particles were evaluated using Monte Carlo simulations with GEANT4. The acceptance was calculated as a function of excitation energy and scattering angle in the center-of-mass system. Because the bending direction of recoil particles in the magnetic field depends on the scattering direction, the acceptance was evaluated separately for particles scattered to the left and to the right.

\section{Results}\label{sec-result}

The cross section were derived using the analysis method described above. The global optical potential (GOP) was determined by considering events with $\left| E_x \right| < 2.5$~MeV in the  cross-section as elastic scattering.
The strength function of ISGMR was obtained through multipole decomposition analysis (MDA), using angular distributions for each $\Delta L$ transition calculated by the GOP and the distorted wave Born approximation (DWBA).

\begin{figure}[th]
    \begin{minipage}[b]{0.5\linewidth}
        \centering
        \includegraphics[keepaspectratio, scale=0.38]{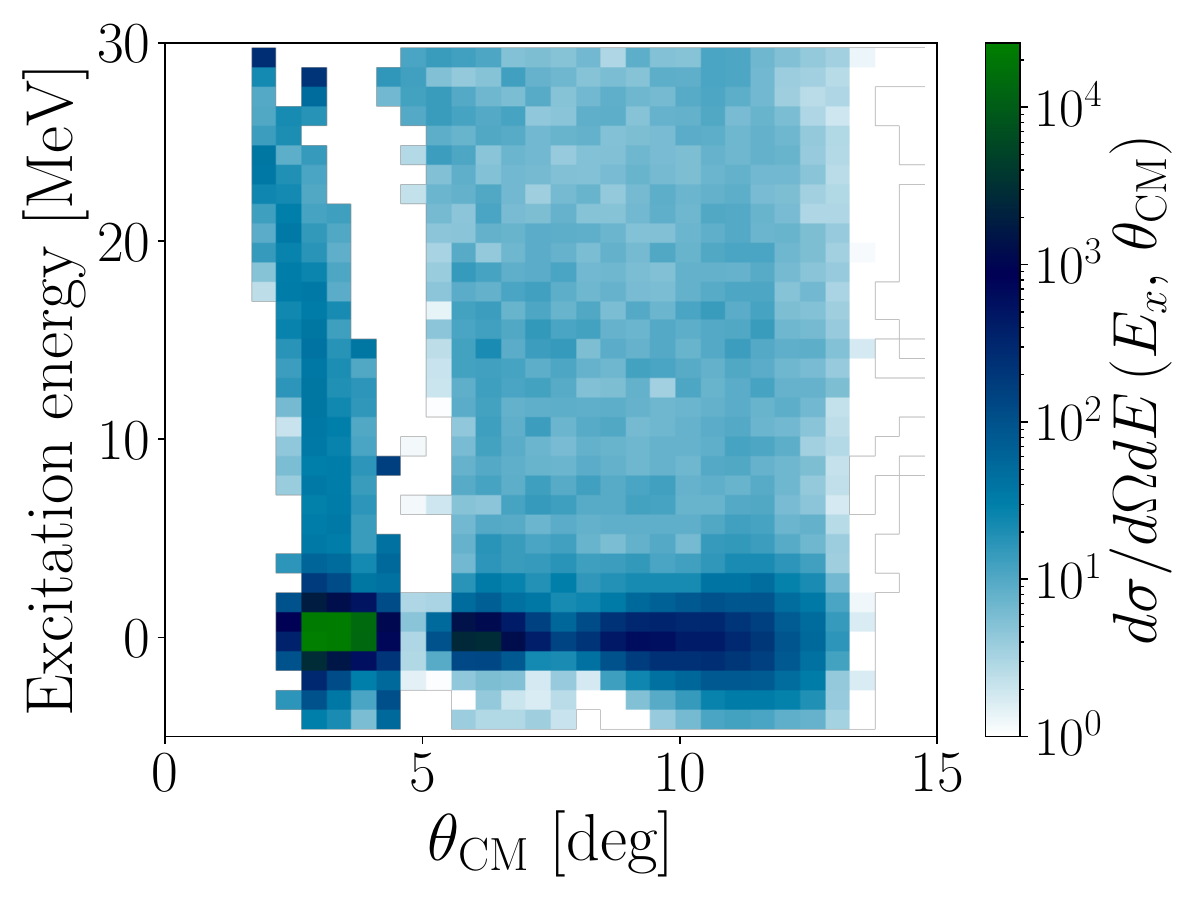}
    \end{minipage}
    \begin{minipage}[b]{0.5\linewidth}
        \centering
        \includegraphics[keepaspectratio, scale=0.38]{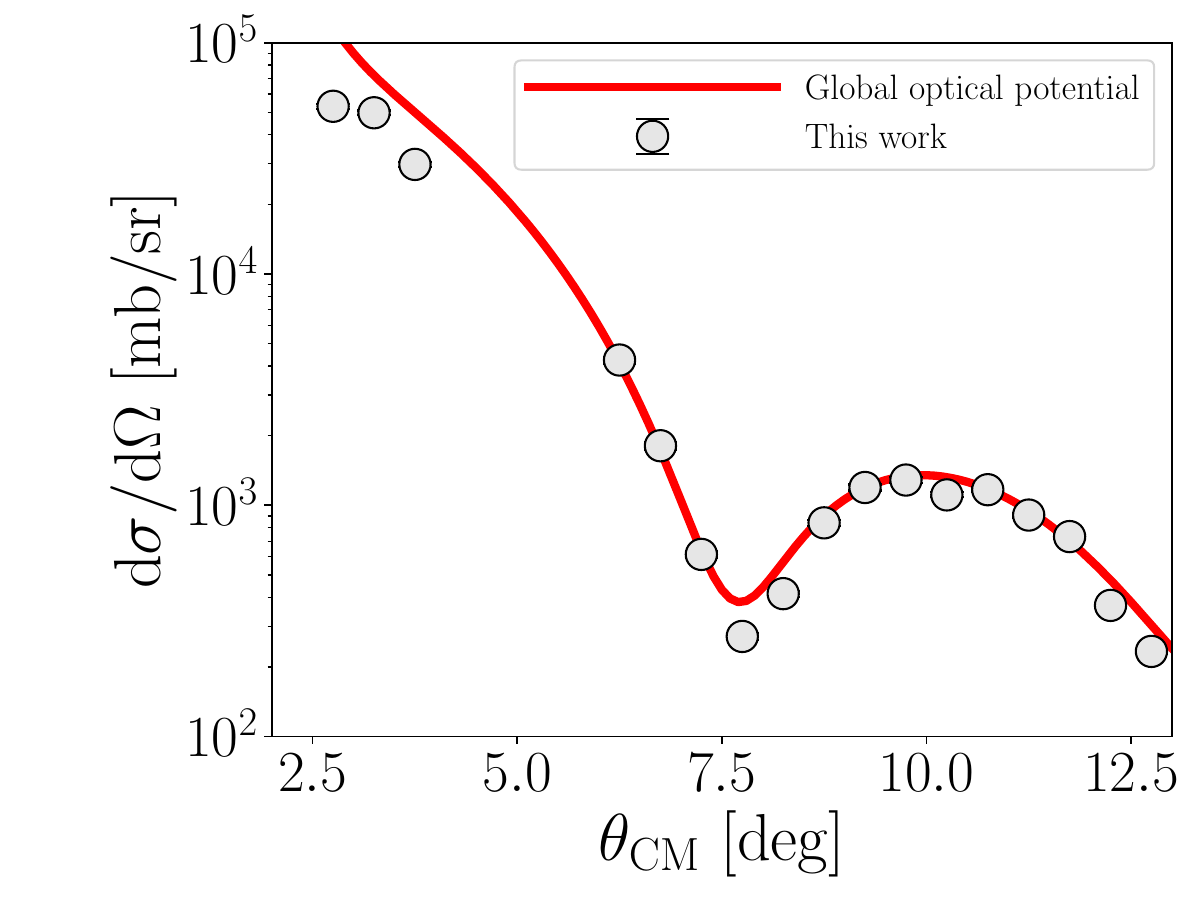}
    \end{minipage}
\caption{Left panel: Double-differential cross section of the $^{86}$Kr$(d,d')$ reaction obtained from this experiment. The cross sections were individually derived for events scattered to the left and right sides and then averaged. Right panel: Differential cross section of elastic scattering, obtained by selecting events within the excitation energy range $|E_x| < 2.5$~MeV. Previous studies have shown that the total cross section for low-lying states below $E_x < 3$~MeV is less than $1$\% of the elastic scattering cross section. The solid curve indicates the calculated cross section by the global optical potential proposed by previous studies.}
\label{fig-dsdode-gop}
\end{figure}

The cross sections were obtained separately for the left side and right side along the beam axis, and their average was taken to derive the  cross section (Figure~\ref{fig-dsdode-gop}). Because the magnetic field causes opposite drift directions on the left and right sides, the angular acceptance differs between them. Accordingly, in regions where only one side was measurable, only the data from that side were used.

Events considered to be elastic scattering are spread over several MeV. According to previous studies, the total cross section to lower-excited states below approximately $E_x \approx 3$~MeV in the $^{86}\mathrm{Kr}(d,d')$ reaction is reported to be less than 1\% of that for elastic scattering \cite{ref-86kr001}. Therefore, we regarded the region within $E_x \pm 2.5$~MeV as elastic scattering. 
The right panel of Figure~\ref{fig-dsdode-gop} shows the differential cross section of elastic scattering. 
The solid line represents the differential cross section calculated using the optical potential determined from the angular distribution of elastic scattering. The optical potential was based on the model proposed by J.~Bojowald \textit{et al.}~\cite{ref-dwba-bojowald}, with only the potential strengths optimized. Details of the optical potential model are summarized in Appendix \ref{ape-dsdode-gop}.
The angular regions below $2.5$~degrees, between $4$ to $6$~degrees, and above $12$~degrees in the angular distribution for elastic scattering were excluded from the analysis. These correspond to events occurring just before the dipole magnet, near the end of the Recoil TPC field cage, or just before or after passing through the SSD, for which accurate acceptance corrections are difficult to estimate.

The strength distribution of isoscalar giant monopole resonance (ISGMR)  was extracted using the multipole decomposition analysis (MDA). 
In the MDA, the cross-section at each excitation energy is represented as a linear combination of angular distributions of ISGMR and other giant resonance components, expressed by the following equation:
\begin{equation}\label{eq-mda-001}
    \frac{d^2\sigma_{\mathrm{meas}}}{dEd\Omega} \bigl( \theta_{\mathrm{CM}},\,E_x\bigr)
    \equiv
    \sum_{\mathrm{L}=0}^{\mathrm{L_{Max}}}
    a_\mathrm{L}\bigl( E_x \bigr)
    \frac{d^2\sigma_\mathrm{DWBA,\,L}}{dEd\Omega} \bigl( \theta_{\mathrm{CM}},\,E_x\bigr)
\end{equation}
where $\sigma_{\mathrm{meas}}$ is the  cross-section obtained from the experiment, and $\sigma_{\mathrm{DWBA,\,L}}$ represents the angular distribution of each multipole component of giant resonances calculated using DWBA. The $a_L$ are the parameters correlated with the strength of each giant resonance

\begin{figure}[th]
    \begin{minipage}[b]{0.5\linewidth}
        \centering
        \includegraphics[keepaspectratio, scale=0.35]{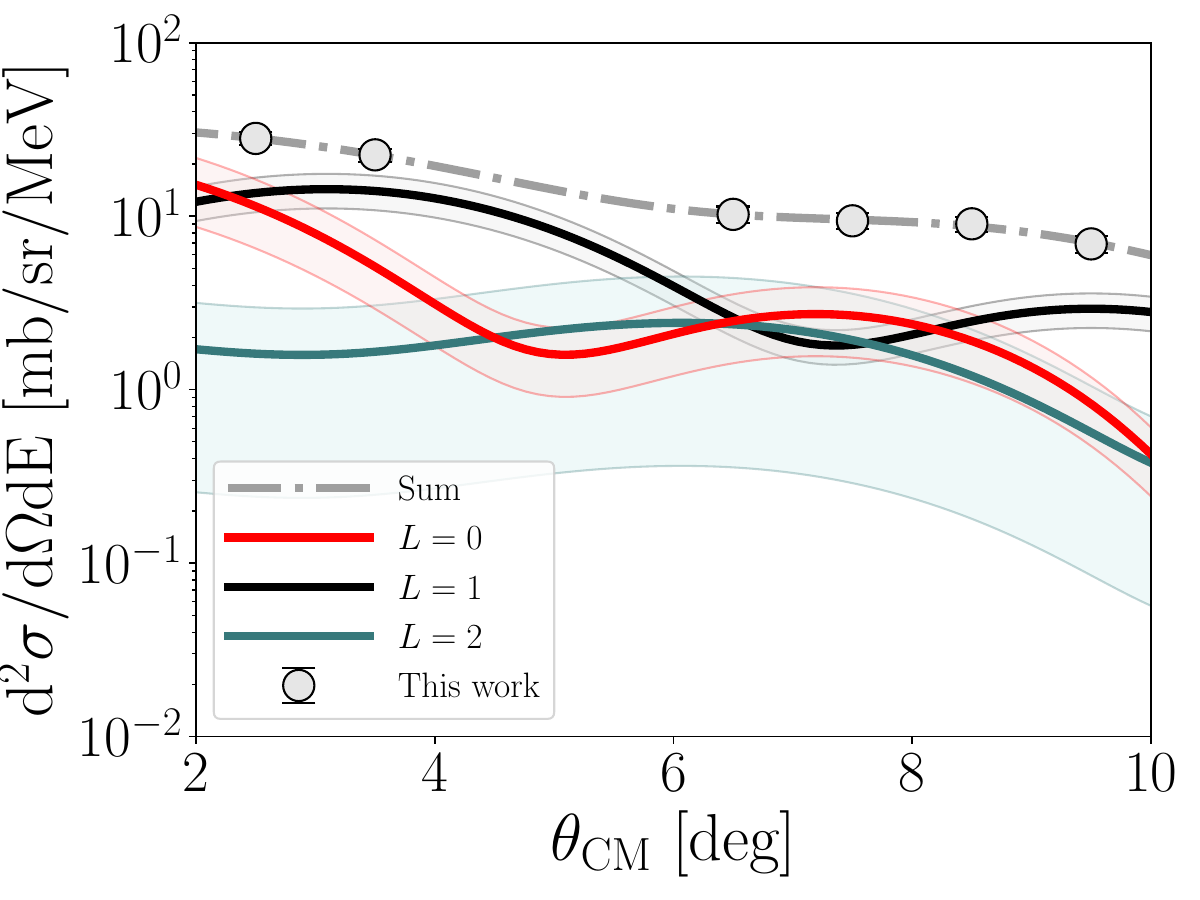}
    \end{minipage}
    \begin{minipage}[b]{0.5\linewidth}
        \centering
        \includegraphics[keepaspectratio, scale=0.35]{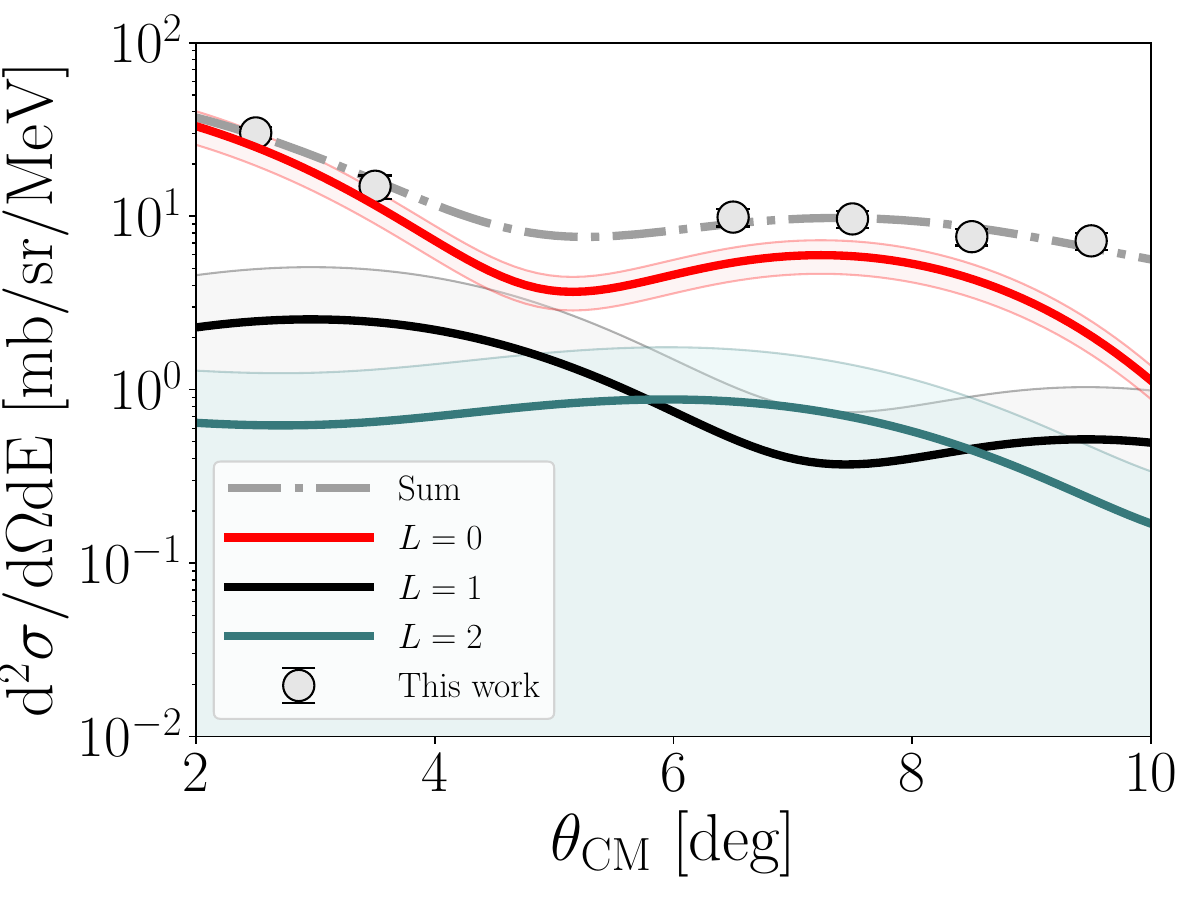}
    \end{minipage}
\caption{Results of the MDA for $^{86}\mathrm{Kr}$. The left figure shows the result for $E_x=11$~MeV, and the right figure shows the result for $E_x=17$~MeV. The solid circles are the differential cross sections obtained from this measurement, and the contributions of ISGMR, ISGQR, and ISGDR are shown as solid blue, orange, and green lines, respectively. The dashed lines are the sums of all multipole components.}
\label{fig-mda-001}
\end{figure}

These parameters were determined by the least-squares method.
The objective function used for the minimization is defined by the following equation:
\begin{equation} \label{eq-mda-002}
    \chi_{\mathrm{MDA}}^2 \left(E_x\right)= 
    \frac{1}{N-L_{\mathrm{Max}}-1}
    \sum_{i=1}^{N}
    \left(
        \frac{ \sigma_{\mathrm{exp}}\left(\theta_i,\,E_x\right) - \sigma_{\mathrm{cal}}\left(\theta_i,\,E_x\right) }
        { \Delta\sigma_{\mathrm{exp}}\left(\theta_i,\,E_x\right) }
    \right)^2
\end{equation}
where the $L_\mathrm{Max}$ is the maximum angular momentum transfer. 
In this study, $L_\mathrm{Max} = 4$ was adopted, and the double-differential cross-section was obtained every $2$ MeV in excitation energy and $1$ degree in scattering angles in the center-of-mass system.
As an example of MDA results, the decomposed angular distribution and differential cross sections at $E_x = 11$ and $17$ MeV are shown in Figure \ref{fig-mda-001}.
The strength with transfer angular momentum $L = 0$ (blue line) corresponds to the ISGMR, $L = 1$ (orange line) to the isoscalar giant dipole resonance (ISGDR), and $L = 2$ (green line) to isoscalar giant quadrupole resonance.
At $E_x = 11$ MeV, ISGMR is comparable to other giant resonance components (ISGQR), but it reaches its maximum at around $E_x=17$ MeV.

\begin{figure}[th]
    \begin{minipage}[b]{0.5\linewidth}
        \centering
        \includegraphics[keepaspectratio, scale=0.35]{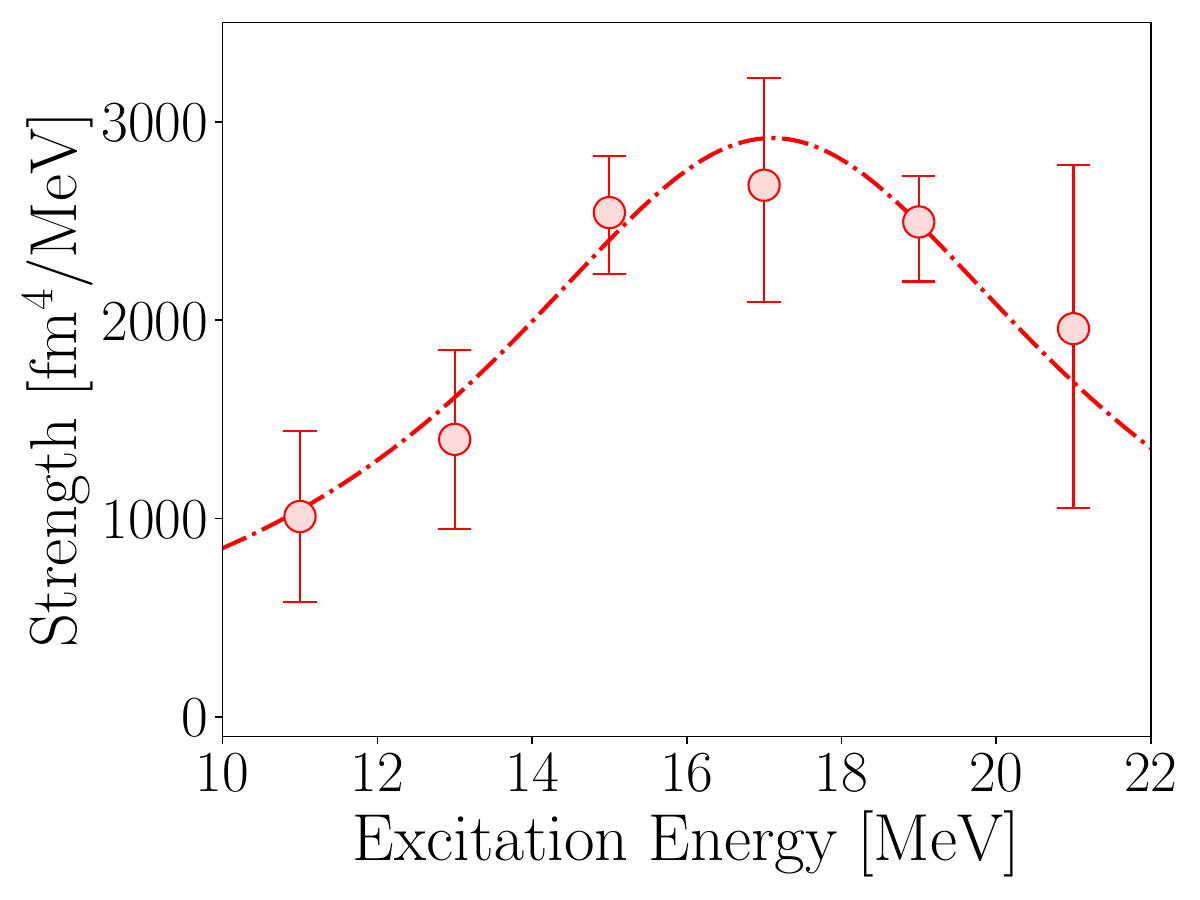}
    \end{minipage}
    \begin{minipage}[b]{0.5\linewidth}
        \centering
        \includegraphics[keepaspectratio, scale=0.35]{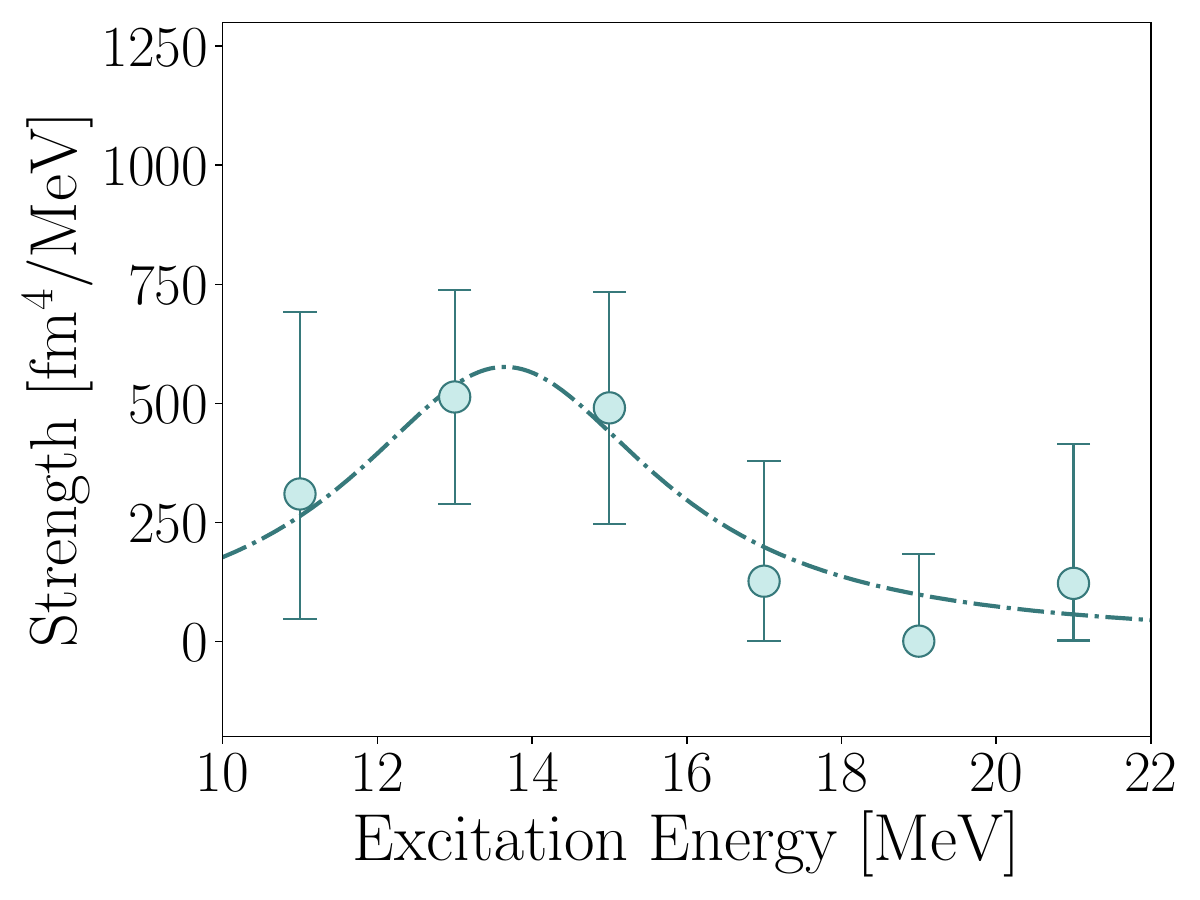}
    \end{minipage}
\caption{Strength functions of the ISGMR and the ISGQR in $^{86}$Kr. The left panel shows the strength function of the isoscalar giant monopole resonance (ISGMR), and the right panel shows that of the isoscalar giant quadrupole resonance (ISGQR). The dashed lines represent the results of fitting with Lorentzian functions.}
\label{fig-gr-001}
\end{figure}

The parameter $a_L(E_x)$ obtained for each excitation energy decomposed by MDA is related to the strength of each transition through the  energy-weighted sum rule (EWSR) by the following equation \cite{ref-isgmr003}:
\begin{equation}
    \begin{split}
        S_{0}(E_x) &= \frac{ 2 \hbar ^2 A }{ m E_x} \left\langle r^2 \right\rangle a_{0}(Ex) \\
        S_{1}(E_x) &= \frac{3\hbar^2 A}{32m\pi E_x} \left( 11 \langle r^4 \rangle - \frac{25}{3} \langle r^2 \rangle^2 - 10 \epsilon \langle r^2 \rangle \right) a_{1}(Ex) \\
        S_{L>2}(E_x) &= \frac{\hbar^2 A}{8\pi m E_x} \lambda(2\lambda + 1)^2  \langle r^{2\lambda-2} \rangle  a_{L}(Ex)
    \end{split}
\end{equation}
where the $A$ is the mass number, the $m$ is the nucleon mass, the $E_x$ is the excitation energy, the $\langle r^n \rangle$ is the n-th moment of the radius in the ground state density, and the $\epsilon$ is denoted as $\frac{\hbar^2}{ 3 m A} \left( \frac{4}{E_{\mathrm{ISGDR}}} + \frac{5}{E_{\mathrm{ISGQR}}} \right)$. 
The strength distributions of the ISGMR and the ISGQR were extracted by these equations (Figure \ref{fig-gr-001}). 
The error bars for each excitation energy were determined by adopting a $68$\% confidence interval using the chi-square test. 

In this study, the deformation effect was neglected, and the strength distributions were fitted with a single Lorentzian function.
The assumption is supported by the relatively small deformation parameter of $^{86}\mathrm{Kr}$ ($\beta_2 = 0.134$) compared to those of nuclei where deformation effects on the ISGMR have been observed, such as $^{148}\mathrm{Sm}$ ($\beta_2 = 0.142$) and $^{94}\mathrm{Mo}$ ($\beta_2 = 0.151$) \cite{ref-sm-isgmr, ref-mo-isgmr001}.
From the fitting results, the centroid energy of the ISGMR was determined to be $E_{0} = 17.1^{+0.7}_{-0.5}$ MeV, and that of the ISGQR was $E_{2} = 13.6^{+1.0}_{-2.9}$ MeV (see Figure \ref{fig-gr-001} and Table \ref{table-mda-001}).
The EWSR values were $174^{+40}_{-43}\%$ for ISGMR and $19^{+21}_{-12}\%$ for ISGQR. 
The centroid energy of ISGMR was determined with a precision of approximately $1$MeV, whereas that of ISGQR had a precision of about $3$MeV.
Similarly, for the EWSR, although a value close to $100$\% was obtained for the ISGMR, the statistical uncertainty was large.
\begin{table}[!h]
    \renewcommand{\arraystretch}{1.5}
    \caption{Summary of fitting parameters for each giant resonances}\label{table-mda-001}
    \centering
    \begin{tabular}{c|c|c|c}
        \hline
        Type &
        Central Energy ($E_{L}$) [MeV] &
        Energy Spread ($\Gamma_{L}$) [MeV] &
        EWSR [\%]\\
        \hline \hline
        ISGMR & $17.1^{+0.7}_{-0.5}$ & $4.5^{+1.7}_{-1.0}$ & $174^{+40}_{-43}$ \\
        ISGQR & $13.6^{+1.0}_{-2.9}$ & $2.4^{+3.3}_{-1.8}$ & $19^{+21}_{-12}$ \\
        \hline
    \end{tabular}
\end{table}

\section{Discussion}\label{sec-discussion}

The centroid energies of the ISGMR and the ISGQR strength functions were compared with those previously obtained in normal kinematics experiments on $N=50$ isotones, and the isospin-dependent term of nuclear matter incompressibility, $K_\tau$, was extracted. As a result, both the systematics of giant resonances and the incompressibility values were found to be consistent, within uncertainties, with those reported in previous studies. Furthermore, this work demonstrates that the ISGMR measurements in proton-rich nuclei near the $N=Z$ region of medium-heavy nuclei are particularly valuable for improving the precision of $K_\tau$ determination as explained below.

    \subsection{Systematics of giant resonances in $N=50$ isotones}\label{sub-discussion-n50-gr}

\begin{figure}[th]
        \begin{minipage}[b]{0.5\linewidth}
            \centering
            \includegraphics[keepaspectratio, scale=0.35]{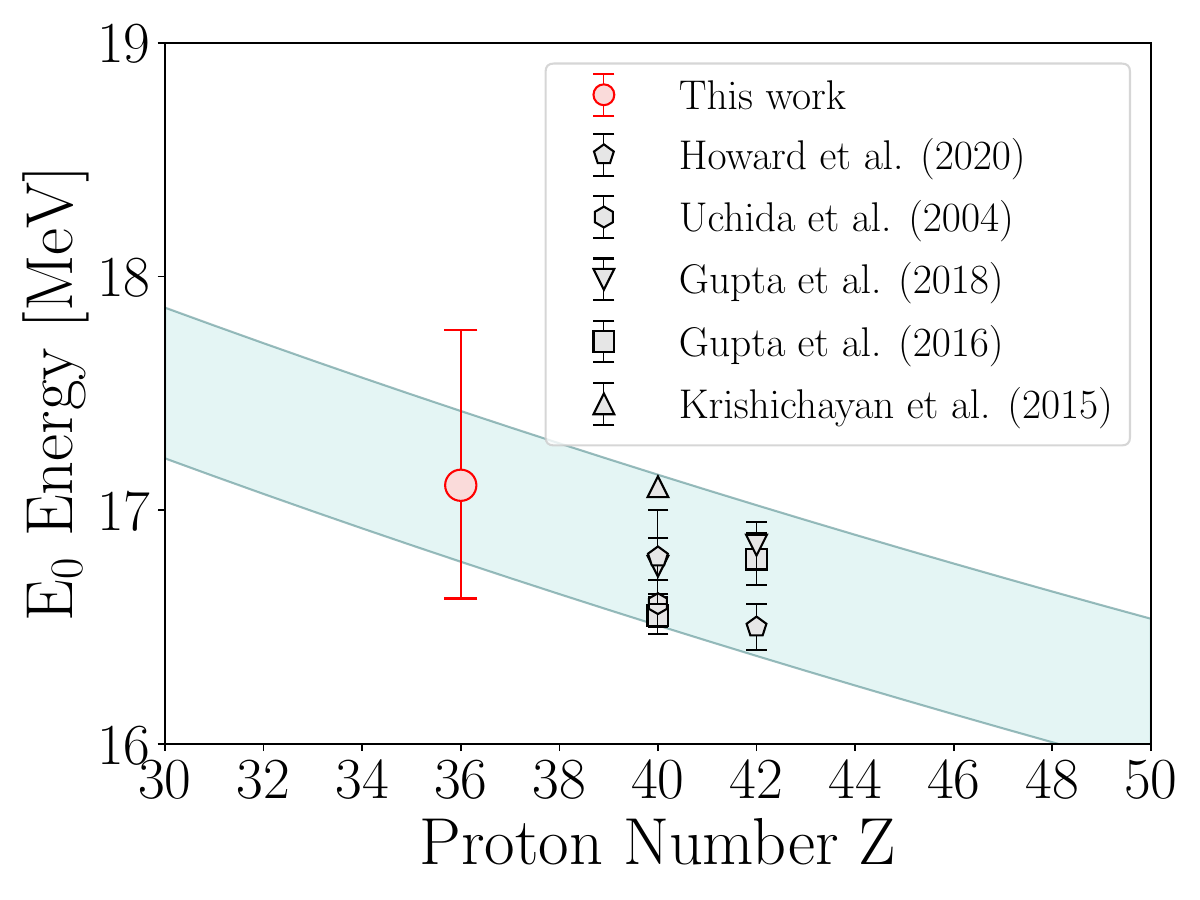}
        \end{minipage}
        \begin{minipage}[b]{0.5\linewidth}
            \centering
            \includegraphics[keepaspectratio, scale=0.35]{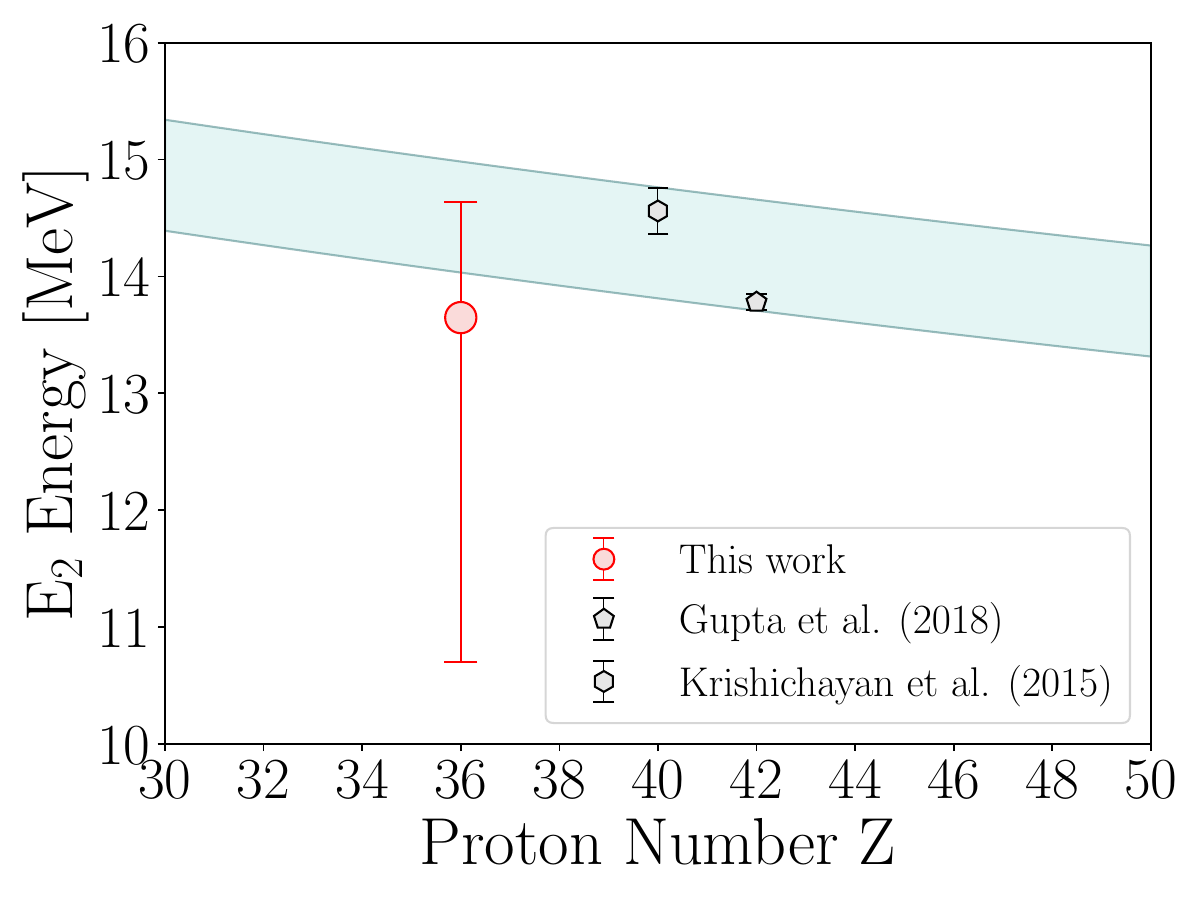}
        \end{minipage}
    \caption{Systematics of the centroid energies of ISGMR and ISGQR strength functions in $N=50$ isotones. The left panel shows the centroid energy of ISGMR, and the right panel shows the centroid energy of ISGQR  as a function of proton number. The red point indicates the results of this work obtained via inverse kinematics, while the black points represent the data from previous studies using normal kinematics. The green hatched bands represent the systematic trends of the giant resonances, scaled as $80A^{-1/3}$ for the ISGMR and $65A^{-1/3}$ for the ISGQR, and shifted to match the range spanned by the existing data.}
    \label{fig-sys-001}
    \end{figure}

    In previous studies, giant resonances of neighboring nuclei have been measured \cite{ref-mo-isgmr001,ref-mo-isgmr002,ref-zr-isgmr001,ref-zr-isgmr002,ref-zr-isgmr003}.
    Among these, $^{90}\mathrm{Zr}$ and $^{92}\mathrm{Mo}$ are isotones with $N=50$ compared to $^{86}\mathrm{Kr}$. 
    Figure \ref{fig-sys-001} shows the comparison of centroid energies derived from the ISGMR measurements of $N=50$ nuclei.
    The black points represent the ISGMR measurement results of $^{90}\mathrm{Zr}$ and $^{92}\mathrm{Mo}$ from previous studies \cite{ref-mo-isgmr001, ref-mo-isgmr002, ref-zr-isgmr001, ref-zr-isgmr002, ref-zr-isgmr003}, and the red point represents the value obtained in this study for $^{86}\mathrm{Kr}$. 
    The green hatched area indicates the extrapolated value based on the macroscopic properties of giant resonances ($80A^{-1/3}$ for ISGMR and $65A^{-1/3}$ for ISGQR), using the maximum and minimum values of energies reported in previous studies. 
    The centroid energies of ISGMR and ISGQR obtained in this study are consistent with this estimated region. 
    Measurement accuracy in this study was approximately three times worse for the ISGMR and ten times worse for the ISGQR.
    
     \begin{table}[th]
        \renewcommand{\arraystretch}{1.5}
        \caption{$E_\mathrm{ISGMR}$ calculated from $E_{\mathrm{centroid}}$, $E_{\mathrm{scaling}}$, and $E_{\mathrm{constrained}}$. Each value is based on the moment ratio ($m_1/m_0$, $\sqrt{m_3/m_1}$ and $\sqrt{m_1/m_{-1}}$). The integration range is $11$-$28$ MeV. The errors are calculated from the parameter errors resulting from the fitting of the strength distribution.}\label{table-ka-001}
        \centering
        \begin{tabular}{c|c|c|c}
            \hline
            Beam &
            $E_\mathrm{scaling}$ ($\sqrt{m_3/m_1}$)         &
            $E_\mathrm{constrained}$ ($\sqrt{m_1/m_{-1}}$)  &
            $E_\mathrm{centroid}$ ($m_1/m_0$)               \\
            \hline \hline
            $^{86}\mathrm{Kr}$                  & 
            $17.3 \pm 1.1_{-0.1}^{+0.0}$  [MeV] & 
            $16.2 \pm 1.0_{-0.1}^{+0.0}$  [MeV] &
            $16.5 \pm 2.0 _{-0.1}^{+0.0}$ [MeV] \\
            \hline
        \end{tabular}
    \end{table}

    \subsection{Nuclear matter incompressibility derived from $N=50$ isotones}\label{sub-discussion-n50-K}

    The nuclear incompressibility $K_{A}$ is derived from the ISGMR strength function by the liquid drop model (Eq.~\ref{ka-dropmodel}) 
    The value of $E_{\mathrm{ISGMR}}$ can be calculated using the following three methods \cite{ref-isgmr001}:
    \begin{equation}
        \begin{split}
            E_{\mathrm{scaling}}      = \sqrt{\frac{m_{3}}{m_{1}}} ,\quad
            E_{\mathrm{constrained}} = \sqrt{\frac{m_{1}}{m_{-1}}} ,\quad
            E_{\mathrm{centroid}}    = \frac{m_{1}}{m_{0}}          
        \end{split}
    \end{equation}
    where
    \begin{equation}\label{eq-moment001}
        m_\mathrm{n} = \int dE_{x} E_{x}^{n} S_L(E_{x})  
    \end{equation}
    It is known that these equations are non-equivalent due to the Schwarz inequality \cite{ref-isgmr003}.
    The results of $E_{\mathrm{ISGMR}}$ for $^{86}\mathrm{Kr}$ calculated using these methods are summarized in Table \ref{table-ka-001}.
    
    In this study, $E_{\mathrm{scaling}}$, obtained via wavefunction scaling, was used to calculate the nuclear incompressibility $K_{A}$. 
    Since $E_{\mathrm{scaling}}$ maintains consistency with the Thomas-Fermi model due to its assumption of constant energy change under density variation, it is suitable as a starting point for discussions on infinite nuclear matter. 

    \begin{figure}[th]
        \begin{minipage}[b]{0.5\linewidth}
            \centering
            \includegraphics[keepaspectratio, scale=0.35]{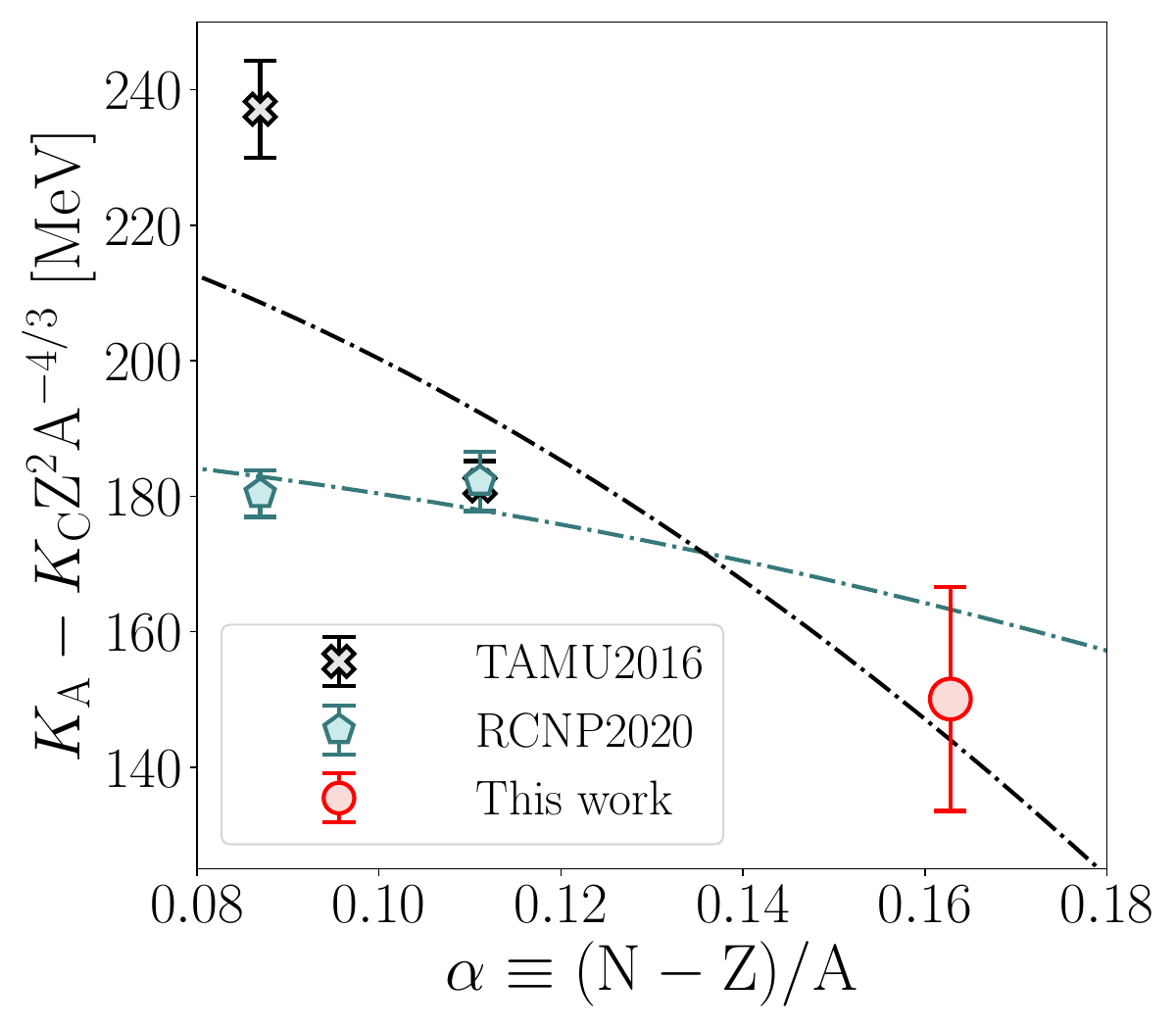}
        \end{minipage}
        \begin{minipage}[b]{0.5\linewidth}
            \centering
            \includegraphics[keepaspectratio, scale=0.3925]{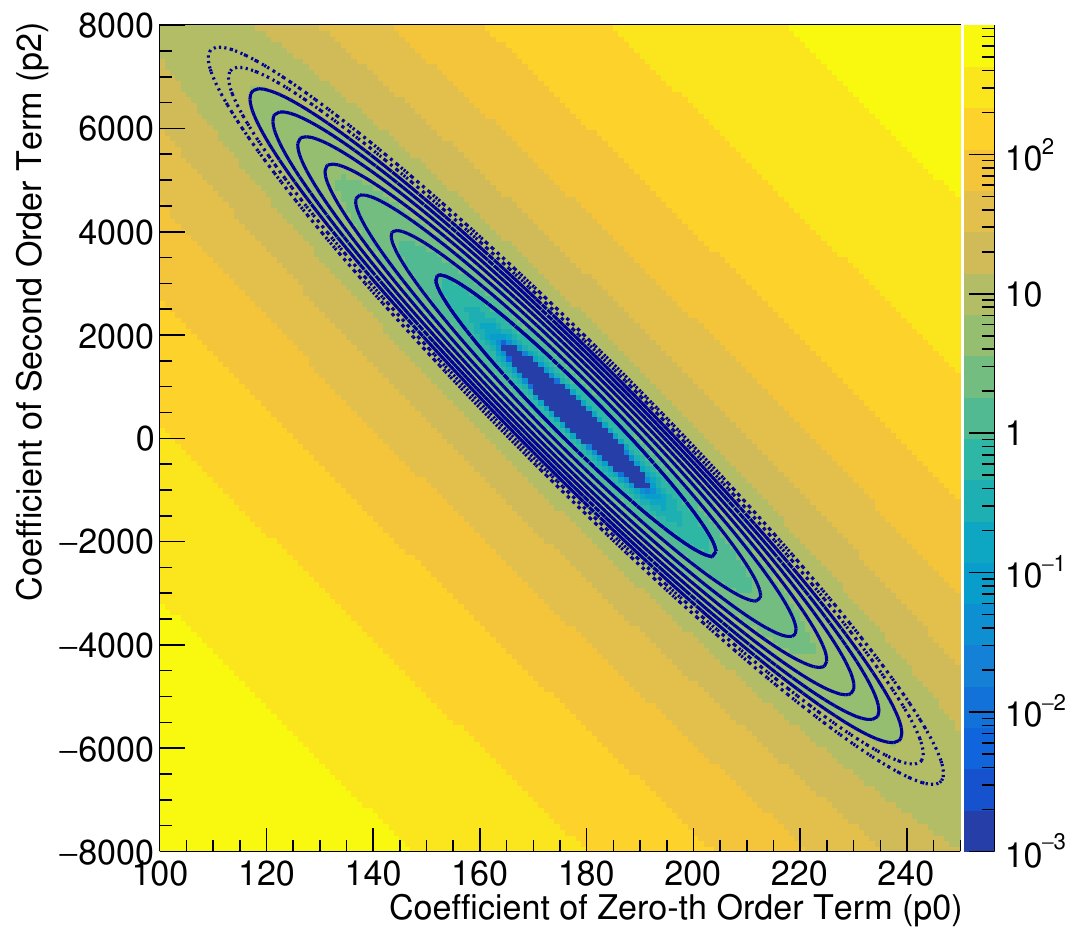}
        \end{minipage}
    \caption{Left: Derivation of $K_\tau$ in $N=50$ isotones. This figure shows the isospin dependence with respect to $K_A - K_\mathrm{C}Z^2A^{-4/3}$. The parameter $K_\tau$ was extracted by fitting the data with a quadratic function based on Eq.~\ref{ka-dropmodel}, and as a result, $K_\tau$ is $-3411 \pm 3245$ MeV for TAMU and $-1035\pm2577$ MeV for RCNP. Right: $\chi^2$ parameter search based on the fit with RCNP data. The figure shows the variation of $\Delta\chi^2 \equiv \chi^2_\mathrm{best,fit} - \chi^2(p_0, p_2)$, with contour lines corresponding to 1, 2, 3,… The innermost contour indicates the 68\% confidence level}
    \label{fig-gr-002}
    \end{figure}

    To derive $K_\tau$, the nuclear incompressibility of $N=50$ isotones was used, and fitting was performed based on the asymmetry dependency. 
    The results are shown in the left panel of Figure \ref{fig-gr-002}. 
    Measurements with $\alpha < 0.12$ are from previous studies conducted at TAMU and RCNP. 
    For data obtained in RCNP, only the most recent values from analyses using multiple methods are considered in this paper. 
    The dashed lines represent the results of deriving $K_\tau$ by fitting TAMU or RCNP data combined with the results of this work. 
    The derived values are $-3411 \pm 3245$ MeV for TAMU and $-1035\pm2577$ MeV for RCNP.
    However, since the TAMU analysis extracted the strength function using Gaussian rather than Lorentzian functions, in the following we focus on a comparison with the RCNP data.

    The $K_\tau$ value obtained in this study has larger uncertainties compared to the previously reported $K_\tau = -550 \pm 100$ MeV. 
    Previous studies derived $K_\tau$ using measurements of six Sn isotopes, while this study derived $K_\tau$ based on three data points for $N=50$ isotone. 
    Therefore, it is as expected that this study does not provide higher accuracy. 
    However, the significance of this study is the consistency between previous studies and this work, which was derived for the first time from the perspective of isotones. 
    Finally, the right panel of Figure \ref{fig-gr-002} shows the confidence intervals of the zeroth- and second-order terms.
    This figure indicates that the error in the zeroth-order term strongly influences the confidence interval of the second-order term. 
    This result highlights the importance of measuring the incompressibility of $N=Z$ nuclei for the precise determination of $K_\tau$.

\section{Conclusion}\label{sec004}

In this study, the ISGMR of $^{86}\mathrm{Kr}$ was measured at HIMAC for the first time using the CAT-M gaseous active target, and we established a method for ISGMR measurement using the active target and successfully derived the absolute value of the ISGMR strength function by applying the MDA to the inverse kinematics experiment with the gaseous active target. 
As a result, the ISGMR energy of $^{86}\mathrm{Kr}$ was determined to be $17.1^{+0.7}_{-0.5}$ MeV. 
This value is consistent with the systematics of the giant resonances in $N=50$ isotones (the ISGMR of $^{90}\mathrm{Zr}$ and $^{92}\mathrm{Mo}$ obtained in previous studies).
In this experiment, a $^{86}\mathrm{Kr}$ beam with a maximum intensity of approximately $1$MHz was used. 
These beam conditions are comparable to those available at facilities providing unstable nuclear beams, such as RIBF. 
Therefore, the measurement method and accuracy achieved in this study are expected to be applicable to future ISGMR measurements of unstable nuclei.

\section*{Acknowledgment}\label{sec005}

This work was supported by JSPS KAKENHI (Grant Number 16H06003) and JST SPRING (Grant Number JPMJSP2114).
This work was performed in part as the Research Project with Heavy Ions at QST-HIMAC with the cooperation of the Accelerator Engineering Corporation (program number 21H445).

\let\doi\relax

\appendix

\section{Modeling of magnetic field map}\label{ape-magnet-model}

\begin{figure}[th]
        \begin{minipage}[b]{0.5\linewidth}
            \centering
            \includegraphics[keepaspectratio, scale=0.35]{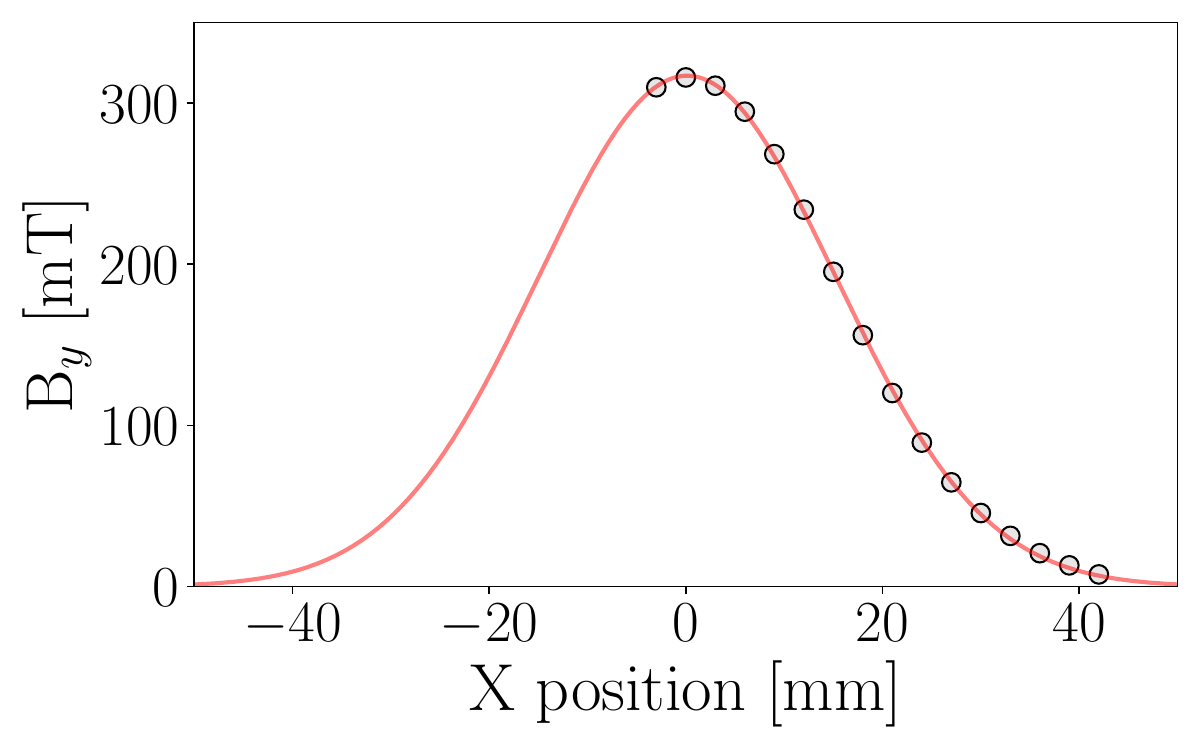}
        \end{minipage}
        \begin{minipage}[b]{0.5\linewidth}
            \centering
            \includegraphics[keepaspectratio, scale=0.35]{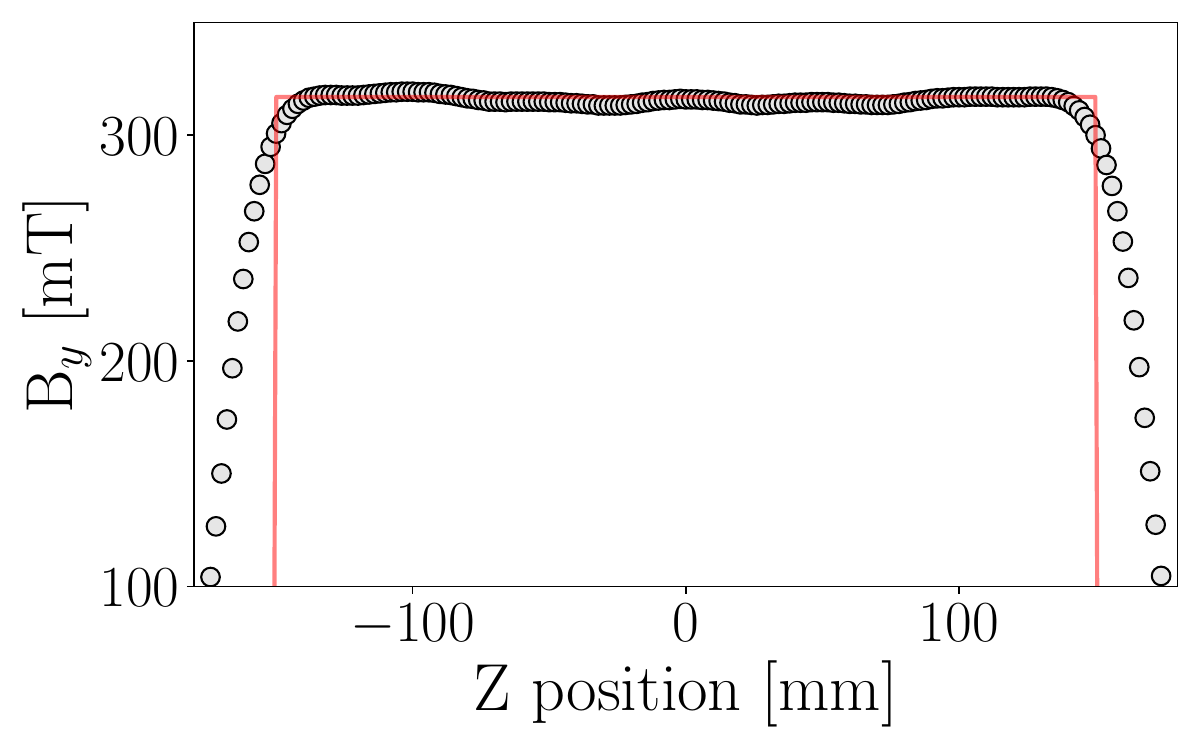}
        \end{minipage}
    \caption{Left: X-direction dependence of the $B_y$ near the center. Since the structure of the dipole magnet and the Recoil TPC is mirror symmetric with respect to the YZ plane at $X = 0$, measurements were conducted only on one side. The field strength attenuates in the X-direction in a manner similar to a Gaussian function and extends up to approximately $40$~mm. Right: Z-direction dependence of the $B_y$ near the center. A sharp drop in the field strength is observed beyond $\pm 150$~mm. The red lines shown in both figures represent the modeled magnetic field adopted in this work.}
    \label{fig-magnet-model-001}
    \end{figure}

    The magnetic field distribution was modeled based on measurements performed by Towa Co., Ltd.
    The dependencies of the $B_y$ ( the component along the electron drift direction) on the X and Z coordinates near the dipole magnet center are shown in Figure \ref{fig-magnet-model-001}.

    Since the dipole magnet and the Recoil TPC are symmetric with respect to the YZ plane, measurements were conducted only on one side.
    The magnetic field strength at the center of the dipole magnet was $B_y = 317$~mT.
    In the X direction, the field leakage extends to approximately $\pm 40$~mm, while in the Z direction, it remains uniform up to around $\pm 135$~mm and then decreases sharply toward $\pm 170$~mm.
    Based on these measurements, the magnetic field distribution was modeled by a Gaussian function in the X-direction combined with a step function in the Z-direction, with parameters determined by fitting (Eq.~\ref{eq-magnet-model}).
    \begin{equation}
     B_y (x,y,z) = 317 \exp \left[-\left(\frac{x - 0.13}{ 15.1 } \right)^2 \right] \times \biggl[ \theta_{\mathrm{step}} \bigl(z - 150 \bigr) -  \theta_{\mathrm{step}} \bigl(z + 150 \bigr) \biggr] \label{eq-magnet-model}
    \end{equation}
    The $B_x$ and $B_z$ components reach their maximum near the edges of the magnet yoke, with magnitudes of approximately $0.02$~T and $0.01$~T, respectively.
    
\begin{figure}[th]
        \begin{minipage}[b]{0.5\linewidth}
            \centering
            \includegraphics[keepaspectratio, scale=0.35]{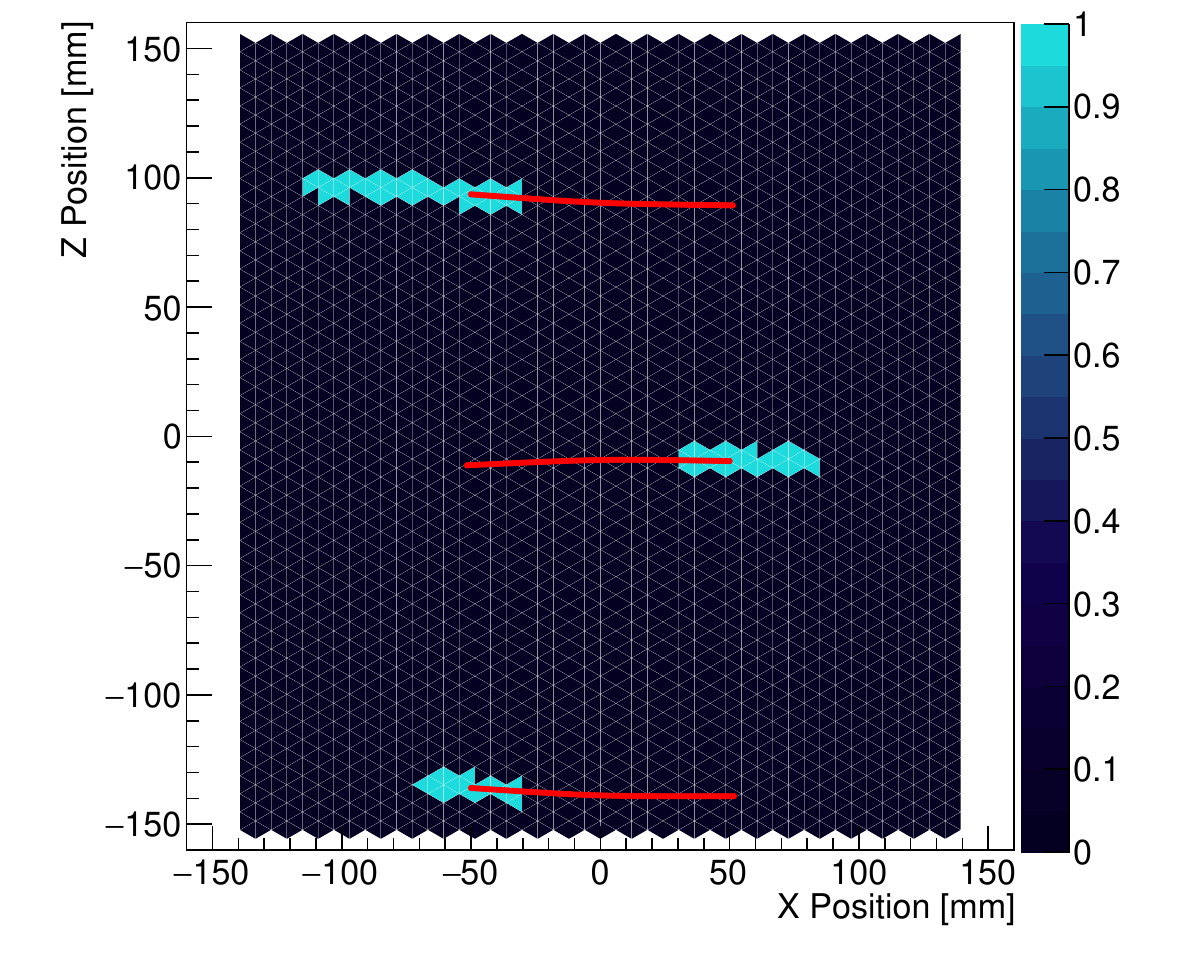}
        \end{minipage}
        \begin{minipage}[b]{0.5\linewidth}
            \centering
            \includegraphics[keepaspectratio, scale=0.35]{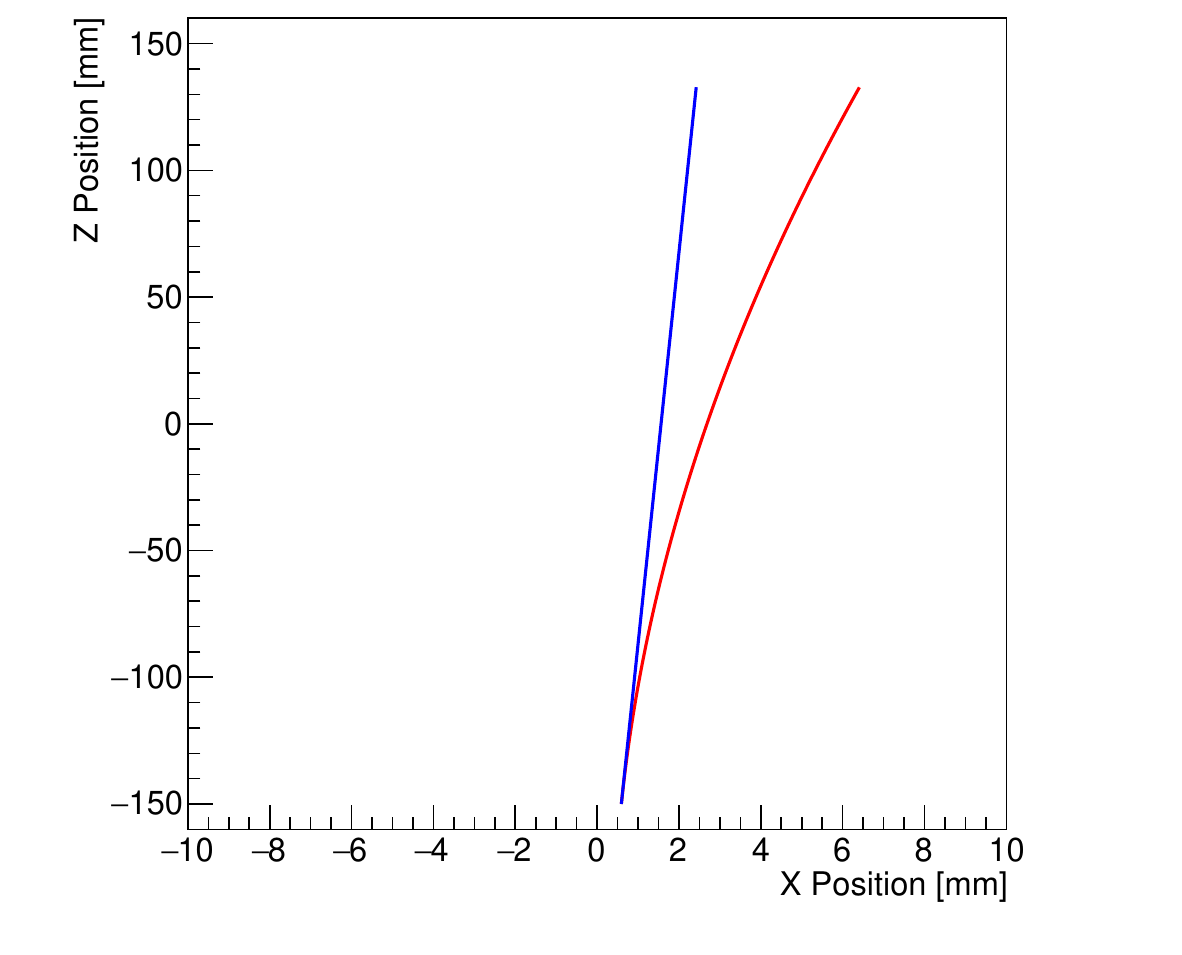}    
        \end{minipage}
    \caption{Left: Hit pattern of recoil particles and the trajectories in the magnetic field. The red lines represent the corrected trajectories in the magnetic field. Right: Trajectories of the beam particle before and after magnetic field correction. The blue line indicates the uncorrected trajectory, while the red line shows the trajectory after correction.}
    \label{fig-track-magnet-001}
    \end{figure}

    The trajectories in the dipole magnet were corrected using the magnetic field model described above (Fig.~\ref{fig-track-magnet-001}).
    The left panel shows the hit patterns of three recoil particles and their corrected trajectories are represented by red lines.
    For recoil particles, the correction range was set to $\pm 50$~mm from the beam axis. The average bending angle in the experimental system was $4.1$~degrees, with a maximum of $6$~degrees.
    This angle does not represent the difference in scattering angles at the vertex, but rather the difference in motion direction between the start and end points of the correction.
    For example, if a recoil particle is scattered to the left side, the angle is calculated as the difference between the motion directions at $+50$~mm and $-50$~mm.
    An example of beam particle trajectories before and after correction is shown in the right panel of Fig.~\ref{fig-track-magnet-001}, where the blue lines represent the uncorrected tracks and the red lines represent the corrected ones.
    For the beam particle, only the central value of the modeled magnetic field was used, and the correction range was set to $\pm 150$~mm.
    The maximum X-direction displacement due to the correction was $6$~mm, while the average and maximum bending angles were $1.5$ and $2.5$~degrees, respectively.

\section{Global optical potential determination} \label{ape-dsdode-gop}
    
\begin{figure}[th]
        \includegraphics[keepaspectratio, scale=0.475]{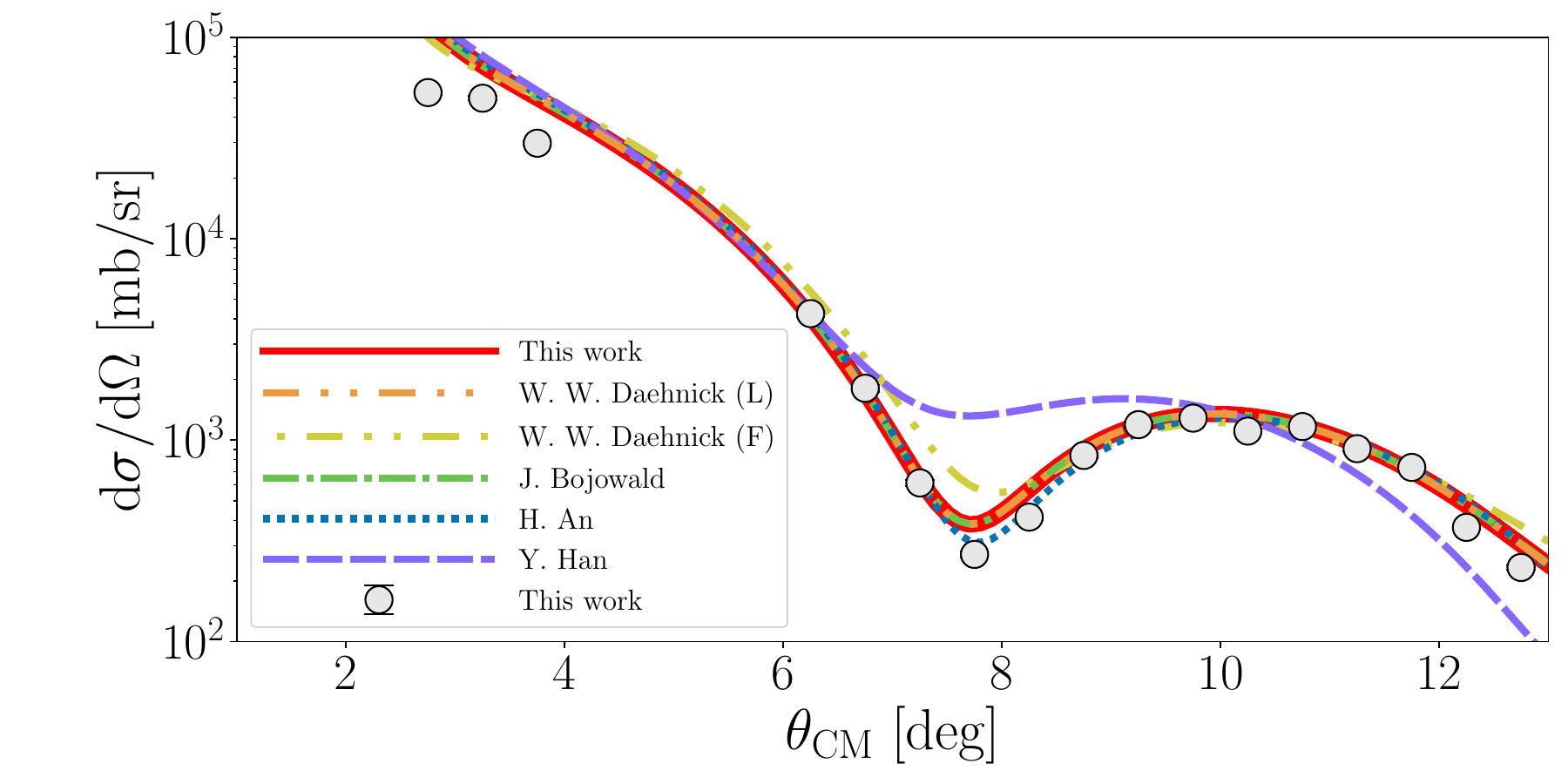}
        \caption{Comparison of global optical potential (GOP) models proposed in previous studies. The data points represent the elastic scattering differential cross sections obtained in this experiment. The red solid line shows the calculated differential cross section using the GOP model proposed by J. Bojowald with optimized potential parameters. The dashed lines represent the differential cross sections calculated using the GOP models with original parameters proposed in previous studies~\cite{ref-dwba-daehnick,ref-dwba-bojowald,ref-dwba-haixia,ref-dwba-han}.}
        \label{fig-dsdode-gop-app}
    \end{figure}

    Several models for the global optical potential of deuterons have been proposed in previous studies~\cite{ref-dwba-daehnick,ref-dwba-bojowald,ref-dwba-haixia,ref-dwba-han}. These optical potentials are expressed as follows:
    \begin{equation} \label{eq-gop-001}
        V = V_{\mathrm{COUL}} + V_{\mathrm{VOL}} + V_{\mathrm{LS}} + i \bigl( W_{\mathrm{VOL}} + W_{\mathrm{SURF}} + W_{\mathrm{LS}} \bigr)
    \end{equation}
    where
    \begin{equation} \label{eq-gop-002}
        \begin{split}
            V_{\mathrm{VOL}}  &= - V_{\mathrm{vol}} f_{\mathrm{v}} \bigl( r,\,R_{\mathrm{v}},\, a_{\mathrm{v}} \bigr) \\
            V_{\mathrm{LS}}   &=   V_{\mathrm{ls}} \biggl( \frac{\hbar}{m_\pi c}\biggr)^2 \frac{1}{r} \bigl( \boldsymbol{L} \cdot \boldsymbol{S} \bigr)
                \frac{d}{dr} f_{\mathrm{ls}} \bigl( r,\,R_{\mathrm{ls}},\,a_{\mathrm{ls}} \bigr) \\
            W_{\mathrm{VOL}}  &= - W_{\mathrm{vol}} f_{\mathrm{wv}} \bigl( r,\,R_{\mathrm{wv}},\,a_{\mathrm{wv}} \bigr) \\
            W_{\mathrm{SURF}}  &= 4 a_{\mathrm{ws}}  W_{\mathrm{surf}} \frac{d}{dr} f_{\mathrm{ws}} \bigl( r,\,R_{\mathrm{ws}},\,a_{\mathrm{ws}} \bigr)\\
            W_{\mathrm{LS}}   &=   W_{\mathrm{ls}} \biggl( \frac{\hbar}{m_\pi c}\biggr)^2 \frac{1}{r} \bigl( \boldsymbol{L} \cdot \boldsymbol{S} \bigr)
                \frac{d}{dr} f_{\mathrm{wls}} \bigl( r,\,R_{\mathrm{wls}},\,a_{\mathrm{wls}} \bigr) \\
        \end{split}
    \end{equation}
    In Eq.~\ref{eq-gop-001}, $V_{\mathrm{X}}$ denotes the real part consisting of Coulomb, volume, and spin-orbit terms. Conversely, $W_{\mathrm{X}}$ represents the imaginary part, composed of volume, surface, and spin-orbit terms. The parameters $V_{\mathrm{x}}$ and $W_{\mathrm{x}}$ in Eq.~\ref{eq-gop-002} denote the strength of each term, while $a_{\mathrm{x}}$, $R_{\mathrm{x}}$, and $f_{\mathrm{x}}$ represent the diffuseness, radius, and Woods--Saxon potential, respectively.
    Figure \ref{fig-dsdode-gop-app} shows the elastic scattering angular distributions calculated from each global optical potential (GOP) model, along with the distribution obtained from the GOP determined using the elastic scattering data from the present experiment.
    
    Among the existing GOP models, only the model proposed by Y.~Han \textit{et al.}~\cite{ref-dwba-han} exhibits a diffraction pattern that deviates significantly from the others. The difference is most likely due to the inclusion of an imaginary spin-orbit term and the use of a different evaluation function, since these are the primary distinctions from the other models. Therefore, in this study, we adopted the model proposed by J.~Bojowald \textit{et al.}~\cite{ref-dwba-bojowald}, which provides the widest applicable energy range, and optimized only the potential strength. The optimized parameters are summarized in Table~\ref{table-gop-001}.
    In addition, angular regions below $2.5$~degrees, between $4$ to $6$~degrees, and above $12$~degrees in the angular distribution for elastic scattering were excluded from the analysis. These correspond to events occurring just before the dipole magnet, near the end of the Recoil TPC field cage, or just before or after passing through the SSD, for which accurate acceptance corrections are difficult to apply.
    
    \begin{table}[th]
         \caption{Summary of the global optical potential parameters for deuteron scattering, including previously proposed models and the optimized parameters obtained in this work.}\label{table-gop-001}
         \centering
          \begin{tabular}{c|cccccccc}
           \hline
           Model &
           $V_\mathrm{vol}$ &
           $R_\mathrm{v}$   &
           $a_\mathrm{v}$   &
           $W_\mathrm{vol}$  &
           $R_\mathrm{wv}$  &
           $a_\mathrm{wv}$  &
           $W_\mathrm{surf}$ &
           $R_\mathrm{ws}$   \\
           \hline \hline
           W. W. Daehnick (L) \cite{ref-dwba-daehnick} & 43.88 & 1.170 & 1.048 & 17.05 & 1.325 & 0.800 & 0.329 & 1.325 \\
           W. W. Daehnick (F) \cite{ref-dwba-daehnick} & 43.88 & 1.170 & 0.956 & 17.83 & 1.235 & 0.789 & 0.344 & 1.235 \\
           J. Bojowald        \cite{ref-dwba-bojowald} & 45.17 & 1.180 & 0.790 & 20.35 & 1.270 & 0.861 & 0.000 & 1.270 \\
           H. An              \cite{ref-dwba-haixia}   & 52.09 & 1.150 & 0.663 & 13.49 & 1.328 & 0.414 & 4.734 & 1.368 \\
           Y. Han             \cite{ref-dwba-han}      & 20.50 & 1.174 & 0.809 & 13.59 & 1.563 & 0.899 &-1.977 & 1.328 \\
           This work                                   & 45.17 & 1.180 & 0.790 & 20.47 & 1.270 & 0.861& 0.000 & 1.270  \\
           \hline
           Model &
           $a_\mathrm{ws}$  &
           $V_\mathrm{ls}$  &
           $R_\mathrm{ls}$  &
           $a_\mathrm{ls}$  &
           $W_\mathrm{ls}$  &
           $R_\mathrm{wls}$ &
           $a_\mathrm{wls}$ &
           $R_\mathrm{c}$   \\
           \hline \hline
           W. W. Daehnick (L) \cite{ref-dwba-daehnick}  & 0.799 & 1.553 & 1.070 & 0.660 & 0.000 & 0.000 & 0.000 & 1.3 \\
           W. W. Daehnick (F) \cite{ref-dwba-daehnick}  & 0.789 & 0.826 & 1.070 & 0.660 & 0.000 & 0.000 & 0.000 & 1.3 \\
           J. Bojowald        \cite{ref-dwba-bojowald}  & 0.861 & 6.000 & 0.948 & 0.948 & 0.000 & 0.000 & 0.000 & 1.3 \\
           H. An              \cite{ref-dwba-haixia}    & 0.805 & 3.557 & 0.972 & 1.011 & 0.000 & 0.000 & 0.000 & 1.3 \\
           Y. Han             \cite{ref-dwba-han}       & 0.664 & 3.703 & 1.234 & 0.813 &-0.206 & 1.234 & 0.813 & 1.3 \\
           This work                                    & 0.861 & 1.270 & 0.948 & 0.948 & 0.000 & 0.000 & 0.000 & 1.3 \\
           \hline
          \end{tabular}
        \end{table}    


\begin{thebibliography}{9}
    \bibitem{ref-eos001} X.~Roca-Maza and N.~Paar, Progress in Particle and Nuclear Physics {\bf 101}, 96-176 (2018). \\ \doi{https://doi.org/10.1016/j.ppnp.2018.04.001}
    
    \bibitem{ref-eos003} F.~J.~Fattoyev et al., Phys. Rev. C {\bf 86}, 025804 (2012). \\
    \doi{https://doi.org/10.1103/PhysRevC.86.025804}    
    
    \bibitem{ref-eos002} H.~Sotani and S.~Ota, Phys. Rev.\ D {\bf 106}, 103005 (2022). \\
    \doi{https://doi.org/10.1103/PhysRevD.106.103005}
    
    \bibitem{ref-isgmr001} J.P.~Blaizot, Physics Reports {\bf 64}, 171-248 (1980). \\
    \doi{https://doi.org/10.1016/0370-1573(80)90001-0}
    
    \bibitem{ref-isgmr005} G.~Colò et al., Phys. Rev. C {\bf 70}, 024307 (2004). \\
    \doi{https://doi.org/10.1103/PhysRevC.70.024307}    
    
    \bibitem{ref-isgmr002} U.~Garg and G.~Colò, Progress in Particle and Nuclear Physics {\bf 101}, 55-95 (2018). \\ \doi{https://doi.org/10.1016/j.ppnp.2018.03.001}
    
    \bibitem{ref-isgmr004} H.~Sagawa et al., Phys. Rev. C {\bf 76}, 034327 (2007). \\ 
    \doi{https://doi.org/10.1103/PhysRevC.76.034327}
    
    \bibitem{ref-maya001} C.E.~Demonchy et al., Nuclear Instruments and Methods in Physics Research Section A: Accelerators, Spectrometers, Detectors and Associated Equipment {\bf 573}, 145-148 (2007). \\ 
    \doi{https://doi.org/10.1016/j.nima.2006.11.025}
    
    \bibitem{ref-maya002} M.~Vandebrouck et al., Phys. Rev. C {\bf 92}, 024316 (2015). \\ 
    \doi{https://doi.org/10.1103/PhysRevC.92.024316}
    
    \bibitem{ref-himac001} Y.~Hirao et al., Nuclear Physics A {\bf 538}, 541-550 (1992). \\ 
    \doi{https://doi.org/10.1016/0375-9474(92)90803-R}
    
    \bibitem{ref-himac002} S.~Sato, T.~Furukawa and K.~Noda, Nuclear Instruments and Methods in Physics Research Section A: Accelerators, Spectrometers, Detectors and Associated Equipment {\bf 574}, 226-231 (2007). \\ 
    \doi{https://doi.org/10.1016/j.nima.2007.01.174}
    
    \bibitem{ref-himac003} K.~Mizushima et al., Nuclear Instruments and Methods in Physics Research Section B: Beam Interactions with Materials and Atoms {\bf 406}, 347-351 (2017). \\ 
    \doi{https://doi.org/10.1016/j.nimb.2017.03.051}
    
    \bibitem{ref-srppac001} S.~Hanai et al., Nuclear Instruments and Methods in Physics Research Section B: Beam Interactions with Materials and Atoms {\bf 541}, 194-196 (2023). \\ 
    \doi{https://doi.org/10.1016/j.nimb.2023.05.016}
    
    \bibitem{ref-srppac002} S.~Hanai et al., [arXiv:2308.12867 [nucl-ex]] \\ 
    \doi{https://doi.org/10.48550/arXiv.2308.12867}
    
    \bibitem{ref-daq006} E.~C.~Pollacco et al., Nuclear Instruments and Methods in Physics Research Section A: Accelerators, Spectrometers, Detectors and Associated Equipment {\bf 887}, 81-93 (2018). \\ 
    \doi{https://doi.org/10.1016/j.nima.2018.01.020}
    
    \bibitem{ref-daq001} H.~Baba et al., Nuclear Instruments and Methods in Physics Research Section A : Accelerators, Spectrometers, Detectors and Associated Equipment {\bf 616}, 65-68 (2010). \\ 
    \doi{https://doi.org/10.1016/j.nima.2010.02.120}
    
    \bibitem{ref-daq002} H.~Baba et al., RIKEN Accel. Prog. Rep. {\bf 45}, 9 (2012).\\
    
    \bibitem{ref-daq005} H.~Baba et al., RIKEN Accel. Prog. Rep. {\bf 53}, 13 (2019).\\
    \doi{https://www.nishina.riken.jp/researcher/APR/APR053/pdf/13.pdf}
    
    \bibitem{ref-daq003} H.~Baba et al., RIKEN Accel. Prog. Rep. {\bf 46}, 213 (2013).\\ 
    
    \bibitem{ref-daq004} H.~Baba et al., RIKEN Accel. Prog. Rep. {\bf 47}, 235 (2014).\\
    \doi{https://www.nishina.riken.jp/researcher/APR/APR047/pdf/235.pdf}
    
    \bibitem{ref-cluster001} R.~R.~Sokal and C.~D.~Michener, University of Kansas Scientific Bulletin, {\bf 28}, 1409-1438 (1958)
    
    \bibitem{ref-srim001} J~F.~Ziegler et al., Nuclear Instruments and Methods in Physics Research Section B: Beam Interactions with Materials and Atoms {\bf 268}, 1818-1823 (2010). \\
    \doi{https://doi.org/10.1016/j.nimb.2010.02.091}
    
    \bibitem{ref-gafieldpp001} R.~Veenhof, Nuclear Instruments and Methods in Physics Research Section A: Accelerators, Spectrometers, Detectors and Associated Equipment {\bf 419}, 726-730 (1998). \\
    \doi{https://doi.org/10.1016/S0168-9002(98)00851-1}
    
    \bibitem{ref-geant4001} S.~Agostinelli et al., Nuclear Instruments and Methods in Physics Research Section A: Accelerators, Spectrometers, Detectors and Associated Equipment, {\bf 506}, 250-303 (2003). \\
    \doi{https://doi.org/10.1016/S0168-9002(03)01368-8}
    
    \bibitem{ref-86kr001} B.~Rosner, E.~J.~Schneid, Nuclear Physics, {\bf 82}, 182-188 (1966). \\ 
    \doi{https://doi.org/10.1016/0029-5582(66)90530-X}
    
    \bibitem{ref-dwba-bojowald} J.~Bojowald et al., Phys. Rev. C {\bf 38}, 1153-1163 (1988). \\ 
    \doi{https://doi.org/10.1103/PhysRevC.38.1153}
    
    \bibitem{ref-isgmr003} M.~N.~Harakeh and A.~Van~der~Woude, Giant Resonances (Oxford University Press, Oxford, 2002). \\ 
    \doi{https://cds.cern.ch/record/579269}
    
    \bibitem{ref-sm-isgmr} M.~Itoh et al., Phys. Rev. C {\bf 68}, 064602 (2003). \\ 
    \doi{https://doi.org/10.1103/PhysRevC.68.064602}
    
    \bibitem{ref-mo-isgmr001} K.B.~Howard et al., Phys. Lett. B {\bf 807}, 135608 (2020). \\ 
    \doi{https://doi.org/10.1016/j.physletb.2020.135608}
    
    \bibitem{ref-mo-isgmr002} Y.~K.~Gupta et al., Phys. Lett. B {\bf 760}, 482-485 (2016). \\ 
    \doi{https://doi.org/10.1016/j.physletb.2016.07.021}
    
    \bibitem{ref-zr-isgmr001} Y.~K.~Gupta et al., Phys. Rev. C {\bf 97}, 064323 (2018). \\ 
    \doi{https://doi.org/10.1103/PhysRevC.97.064323}
    
    \bibitem{ref-zr-isgmr002} Krishichayan et al., Phys. Rev. C {\bf 92}, 044323 (2015). \\ 
    \doi{https://doi.org/10.1103/PhysRevC.92.044323}
    
    \bibitem{ref-zr-isgmr003} M.~Uchida et al., Phys. Rev. C {\bf 69}, 051301 (2004). \\ 
    \doi{https://doi.org/10.1103/PhysRevC.69.051301} 
    
    \bibitem{ref-dwba-daehnick} W.~W.~Daehnick, J.~D.~Childs and Z.~Vrcelj, Phys. Rev. C {\bf 21}, 2253-2274 (1980). \\ \doi{https://doi.org/10.1103/PhysRevC.21.2253}
    
    \bibitem{ref-dwba-haixia} H.~An and C.~Cai, Phys. Rev. C {\bf 74}, 054605 (2006). \\ 
    \doi{https://doi.org/10.1103/PhysRevC.73.054605}
    
    \bibitem{ref-dwba-han} Y.~Han, Y.~Shi and Q.~Shen, Phys. Rev. C {\bf 74}, 044615 (2006). \\ 
    \doi{https://doi.org/10.1103/PhysRevC.74.044615}  
\end{thebibliography}
\end{document}